\magnification=\magstep1
\def\firstpage{1}
\pageno=\firstpage
\font\fiverm=cmr5
\font\sevenrm=cmr7
\font\sevenbf=cmbx7
\font\eightrm=cmr8
\font\eightbf=cmbx8
\font\ninerm=cmr9
\font\ninebf=cmbx9
\font\tenbf=cmbx10
\font\twelvebf=cmbx12

\font\magnineeufm=eufm9 scaled\magstep1

%
%
\newskip\ttglue
\font\fiverm=cmr5
\font\fivei=cmmi5
\font\fivesy=cmsy5
\font\fivebf=cmbx5
\font\sixrm=cmr6
\font\sixi=cmmi6
\font\sixsy=cmsy6
\font\sixbf=cmbx6
\font\sevenrm=cmr7
\font\eightrm=cmr8
\font\eighti=cmmi8
\font\eightsy=cmsy8
\font\eightit=cmti8
\font\eightsl=cmsl8
\font\eighttt=cmtt8
\font\eightbf=cmbx8
\font\ninerm=cmr9
\font\ninei=cmmi9
\font\ninesy=cmsy9
\font\nineit=cmti9
\font\ninesl=cmsl9
\font\ninett=cmtt9
\font\ninebf=cmbx9
\font\twelverm=cmr12
\font\twelvei=cmmi12
\font\twelvesy=cmsy12
\font\twelveit=cmti12
\font\twelvesl=cmsl12
\font\twelvett=cmtt12
\font\twelvebf=cmbx12


\def\eightpoint{\def\rm{\fam0\eightrm}  
  \textfont0=\eightrm \scriptfont0=\sixrm \scriptscriptfont0=\fiverm
  \textfont1=\eighti  \scriptfont1=\sixi  \scriptscriptfont1=\fivei
  \textfont2=\eightsy  \scriptfont2=\sixsy  \scriptscriptfont2=\fivesy
  \textfont3=\tenex  \scriptfont3=\tenex  \scriptscriptfont3=\tenex
  \textfont\itfam=\eightit  \def\it{\fam\itfam\eightit}
  \textfont\slfam=\eightsl  \def\sl{\fam\slfam\eightsl}
  \textfont\ttfam=\eighttt  \def\tt{\fam\ttfam\eighttt}
  \textfont\bffam=\eightbf  \scriptfont\bffam=\sixbf
    \scriptscriptfont\bffam=\fivebf  \def\bf{\fam\bffam\eightbf}
  \tt  \ttglue=.5em plus.25em minus.15em
  \normalbaselineskip=9pt
  \setbox\strutbox=\hbox{\vrule height7pt depth2pt width0pt}
  \let\sc=\sixrm  \let\big=\eightbig \normalbaselines\rm}

\def\eightbig#1{{\hbox{$\textfont0=\ninerm\textfont2=\ninesy
        \left#1\vbox to6.5pt{}\right.$}}}


\def\ninepoint{\def\rm{\fam0\ninerm}  
  \textfont0=\ninerm \scriptfont0=\sixrm \scriptscriptfont0=\fiverm
  \textfont1=\ninei  \scriptfont1=\sixi  \scriptscriptfont1=\fivei
  \textfont2=\ninesy  \scriptfont2=\sixsy  \scriptscriptfont2=\fivesy
  \textfont3=\tenex  \scriptfont3=\tenex  \scriptscriptfont3=\tenex
  \textfont\itfam=\nineit  \def\it{\fam\itfam\nineit}
  \textfont\slfam=\ninesl  \def\sl{\fam\slfam\ninesl}
  \textfont\ttfam=\ninett  \def\tt{\fam\ttfam\ninett}
  \textfont\bffam=\ninebf  \scriptfont\bffam=\sixbf
    \scriptscriptfont\bffam=\fivebf  \def\bf{\fam\bffam\ninebf}
  \tt  \ttglue=.5em plus.25em minus.15em
  \normalbaselineskip=11pt
  \setbox\strutbox=\hbox{\vrule height8pt depth3pt width0pt}
  \let\sc=\sevenrm  \let\big=\ninebig \normalbaselines\rm}

\def\ninebig#1{{\hbox{$\textfont0=\tenrm\textfont2=\tensy
        \left#1\vbox to7.25pt{}\right.$}}}


\def\twelvepoint{\def\rm{\fam0\twelverm}  
  \textfont0=\twelverm \scriptfont0=\eightrm \scriptscriptfont0=\sixrm
  \textfont1=\twelvei  \scriptfont1=\eighti  \scriptscriptfont1=\sixi
  \textfont2=\twelvesy  \scriptfont2=\eightsy  \scriptscriptfont2=\sixsy
  \textfont3=\tenex  \scriptfont3=\tenex  \scriptscriptfont3=\tenex
  \textfont\itfam=\twelveit  \def\it{\fam\itfam\twelveit}
  \textfont\slfam=\twelvesl  \def\sl{\fam\slfam\twelvesl}
  \textfont\ttfam=\twelvett  \def\tt{\fam\ttfam\twelvett}
  \textfont\bffam=\twelvebf  \scriptfont\bffam=\eightbf
    \scriptscriptfont\bffam=\sixbf  \def\bf{\fam\bffam\twelvebf}
  \tt  \ttglue=.5em plus.25em minus.15em
  \normalbaselineskip=11pt
  \setbox\strutbox=\hbox{\vrule height8pt depth3pt width0pt}
  \let\sc=\sevenrm  \let\big=\twelvebig \normalbaselines\rm}

\def\twelvebig#1{{\hbox{$\textfont0=\tenrm\textfont2=\tensy
        \left#1\vbox to7.25pt{}\right.$}}}
\catcode`\@=11
%

\def\undefine#1{\let#1\undefined}
\def\newsymbol#1#2#3#4#5{\let\next@\relax
 \ifnum#2=\@ne\let\next@\msafam@\else
 \ifnum#2=\tw@\let\next@\msbfam@\fi\fi
 \mathchardef#1="#3\next@#4#5}
\def\mathhexbox@#1#2#3{\relax
 \ifmmode\mathpalette{}{\m@th\mathchar"#1#2#3}%
 \else\leavevmode\hbox{$\m@th\mathchar"#1#2#3$}\fi}
\def\hexnumber@#1{\ifcase#1 0\or 1\or 2\or 3\or 4\or 5\or 6\or 7\or 8\or
 9\or A\or B\or C\or D\or E\or F\fi}

\font\tenmsa=msam10
\font\sevenmsa=msam7
\font\fivemsa=msam5
\newfam\msafam
\textfont\msafam=\tenmsa
\scriptfont\msafam=\sevenmsa
\scriptscriptfont\msafam=\fivemsa
\edef\msafam@{\hexnumber@\msafam}
\mathchardef\dabar@"0\msafam@39
\def\dashrightarrow{\mathrel{\dabar@\dabar@\mathchar"0\msafam@4B}}
\def\dashleftarrow{\mathrel{\mathchar"0\msafam@4C\dabar@\dabar@}}

\def\ulcorner{\delimiter"4\msafam@70\msafam@70 }
\def\urcorner{\delimiter"5\msafam@71\msafam@71 }
\def\llcorner{\delimiter"4\msafam@78\msafam@78 }
\def\lrcorner{\delimiter"5\msafam@79\msafam@79 }
\def\yen{{\mathhexbox@\msafam@55}}
\def\checkmark{{\mathhexbox@\msafam@58}}
\def\circledR{{\mathhexbox@\msafam@72}}
\def\maltese{{\mathhexbox@\msafam@7A}}

\font\tenmsb=msbm10
\font\sevenmsb=msbm7
\font\fivemsb=msbm5
\newfam\msbfam
\textfont\msbfam=\tenmsb
\scriptfont\msbfam=\sevenmsb
\scriptscriptfont\msbfam=\fivemsb
\edef\msbfam@{\hexnumber@\msbfam}
\def\Bbb#1{{\fam\msbfam\relax#1}}
\def\widehat#1{\setbox\z@\hbox{$\m@th#1$}%
 \ifdim\wd\z@>\tw@ em\mathaccent"0\msbfam@5B{#1}%
 \else\mathaccent"0362{#1}\fi}
\def\widetilde#1{\setbox\z@\hbox{$\m@th#1$}%
 \ifdim\wd\z@>\tw@ em\mathaccent"0\msbfam@5D{#1}%
 \else\mathaccent"0365{#1}\fi}
\font\teneufm=eufm10
\font\seveneufm=eufm7
\font\fiveeufm=eufm5
\newfam\eufmfam
\textfont\eufmfam=\teneufm
\scriptfont\eufmfam=\seveneufm
\scriptscriptfont\eufmfam=\fiveeufm

\catcode`\@=11
\newsymbol\boxdot 1200
\newsymbol\boxplus 1201
\newsymbol\boxtimes 1202
\newsymbol\square 1003
\newsymbol\blacksquare 1004
\newsymbol\centerdot 1205
\newsymbol\lozenge 1006
\newsymbol\blacklozenge 1007
\newsymbol\circlearrowright 1308
\newsymbol\circlearrowleft 1309
\undefine\rightleftharpoons
\newsymbol\rightleftharpoons 130A
\newsymbol\leftrightharpoons 130B
\newsymbol\boxminus 120C
\newsymbol\Vdash 130D
\newsymbol\Vvdash 130E
\newsymbol\vDash 130F
\newsymbol\twoheadrightarrow 1310
\newsymbol\twoheadleftarrow 1311
\newsymbol\leftleftarrows 1312
\newsymbol\rightrightarrows 1313
\newsymbol\upuparrows 1314
\newsymbol\downdownarrows 1315
\newsymbol\upharpoonright 1316
 
\newsymbol\downharpoonright 1317
\newsymbol\upharpoonleft 1318
\newsymbol\downharpoonleft 1319
\newsymbol\rightarrowtail 131A
\newsymbol\leftarrowtail 131B
\newsymbol\leftrightarrows 131C
\newsymbol\rightleftarrows 131D
\newsymbol\Lsh 131E
\newsymbol\Rsh 131F
\newsymbol\rightsquigarrow 1320
\newsymbol\leftrightsquigarrow 1321
\newsymbol\looparrowleft 1322
\newsymbol\looparrowright 1323
\newsymbol\circeq 1324
\newsymbol\succsim 1325
\newsymbol\gtrsim 1326
\newsymbol\gtrapprox 1327
\newsymbol\multimap 1328
\newsymbol\therefore 1329
\newsymbol\because 132A
\newsymbol\doteqdot 132B
 
\newsymbol\triangleq 132C
\newsymbol\precsim 132D
\newsymbol\lesssim 132E
\newsymbol\lessapprox 132F
\newsymbol\eqslantless 1330
\newsymbol\eqslantgtr 1331
\newsymbol\curlyeqprec 1332
\newsymbol\curlyeqsucc 1333
\newsymbol\preccurlyeq 1334
\newsymbol\leqq 1335
\newsymbol\leqslant 1336
\newsymbol\lessgtr 1337
\newsymbol\backprime 1038
\newsymbol\risingdotseq 133A
\newsymbol\fallingdotseq 133B
\newsymbol\succcurlyeq 133C
\newsymbol\geqq 133D
\newsymbol\geqslant 133E
\newsymbol\gtrless 133F
\newsymbol\sqsubset 1340
\newsymbol\sqsupset 1341
\newsymbol\vartriangleright 1342
\newsymbol\vartriangleleft 1343
\newsymbol\trianglerighteq 1344
\newsymbol\trianglelefteq 1345
\newsymbol\bigstar 1046
\newsymbol\between 1347
\newsymbol\blacktriangledown 1048
\newsymbol\blacktriangleright 1349
\newsymbol\blacktriangleleft 134A
\newsymbol\vartriangle 134D
\newsymbol\blacktriangle 104E
\newsymbol\triangledown 104F
\newsymbol\eqcirc 1350
\newsymbol\lesseqgtr 1351
\newsymbol\gtreqless 1352
\newsymbol\lesseqqgtr 1353
\newsymbol\gtreqqless 1354
\newsymbol\Rrightarrow 1356
\newsymbol\Lleftarrow 1357
\newsymbol\veebar 1259
\newsymbol\barwedge 125A
\newsymbol\doublebarwedge 125B
\undefine\angle
\newsymbol\angle 105C
\newsymbol\measuredangle 105D
\newsymbol\sphericalangle 105E
\newsymbol\varpropto 135F
\newsymbol\smallsmile 1360
\newsymbol\smallfrown 1361
\newsymbol\Subset 1362
\newsymbol\Supset 1363
\newsymbol\Cup 1264
 
\newsymbol\Cap 1265
 
\newsymbol\curlywedge 1266
\newsymbol\curlyvee 1267
\newsymbol\leftthreetimes 1268
\newsymbol\rightthreetimes 1269
\newsymbol\subseteqq 136A
\newsymbol\supseteqq 136B
\newsymbol\bumpeq 136C
\newsymbol\Bumpeq 136D
\newsymbol\lll 136E
 
\newsymbol\ggg 136F
 
\newsymbol\circledS 1073
\newsymbol\pitchfork 1374
\newsymbol\dotplus 1275
\newsymbol\backsim 1376
\newsymbol\backsimeq 1377
\newsymbol\complement 107B
\newsymbol\intercal 127C
\newsymbol\circledcirc 127D
\newsymbol\circledast 127E
\newsymbol\circleddash 127F
\newsymbol\lvertneqq 2300
\newsymbol\gvertneqq 2301
\newsymbol\nleq 2302
\newsymbol\ngeq 2303
\newsymbol\nless 2304
\newsymbol\ngtr 2305
\newsymbol\nprec 2306
\newsymbol\nsucc 2307
\newsymbol\lneqq 2308
\newsymbol\gneqq 2309
\newsymbol\nleqslant 230A
\newsymbol\ngeqslant 230B
\newsymbol\lneq 230C
\newsymbol\gneq 230D
\newsymbol\npreceq 230E
\newsymbol\nsucceq 230F
\newsymbol\precnsim 2310
\newsymbol\succnsim 2311
\newsymbol\lnsim 2312
\newsymbol\gnsim 2313
\newsymbol\nleqq 2314
\newsymbol\ngeqq 2315
\newsymbol\precneqq 2316
\newsymbol\succneqq 2317
\newsymbol\precnapprox 2318
\newsymbol\succnapprox 2319
\newsymbol\lnapprox 231A
\newsymbol\gnapprox 231B
\newsymbol\nsim 231C
\newsymbol\ncong 231D
\newsymbol\diagup 201E
\newsymbol\diagdown 201F
\newsymbol\varsubsetneq 2320
\newsymbol\varsupsetneq 2321
\newsymbol\nsubseteqq 2322
\newsymbol\nsupseteqq 2323
\newsymbol\subsetneqq 2324
\newsymbol\supsetneqq 2325
\newsymbol\varsubsetneqq 2326
\newsymbol\varsupsetneqq 2327
\newsymbol\subsetneq 2328
\newsymbol\supsetneq 2329
\newsymbol\nsubseteq 232A
\newsymbol\nsupseteq 232B
\newsymbol\nparallel 232C
\newsymbol\nmid 232D
\newsymbol\nshortmid 232E
\newsymbol\nshortparallel 232F
\newsymbol\nvdash 2330
\newsymbol\nVdash 2331
\newsymbol\nvDash 2332
\newsymbol\nVDash 2333
\newsymbol\ntrianglerighteq 2334
\newsymbol\ntrianglelefteq 2335
\newsymbol\ntriangleleft 2336
\newsymbol\ntriangleright 2337
\newsymbol\nleftarrow 2338
\newsymbol\nrightarrow 2339
\newsymbol\nLeftarrow 233A
\newsymbol\nRightarrow 233B
\newsymbol\nLeftrightarrow 233C
\newsymbol\nleftrightarrow 233D
\newsymbol\divideontimes 223E
\newsymbol\varnothing 203F
\newsymbol\nexists 2040
\newsymbol\Finv 2060
\newsymbol\Game 2061
\newsymbol\mho 2066
\newsymbol\eth 2067
\newsymbol\eqsim 2368
\newsymbol\beth 2069
\newsymbol\gimel 206A
\newsymbol\daleth 206B
\newsymbol\lessdot 236C
\newsymbol\gtrdot 236D
\newsymbol\ltimes 226E
\newsymbol\rtimes 226F
\newsymbol\shortmid 2370
\newsymbol\shortparallel 2371
\newsymbol\smallsetminus 2272
\newsymbol\thicksim 2373
\newsymbol\thickapprox 2374
\newsymbol\approxeq 2375
\newsymbol\succapprox 2376
\newsymbol\precapprox 2377
\newsymbol\curvearrowleft 2378
\newsymbol\curvearrowright 2379
\newsymbol\digamma 207A
\newsymbol\varkappa 207B
\newsymbol\Bbbk 207C
\newsymbol\hslash 207D
\undefine\hbar
\newsymbol\hbar 207E
\newsymbol\backepsilon 237F

%
\newcount\marknumber	\marknumber=1
\newcount\countdp \newcount\countwd \newcount\countht 
%
%
\ifx\pdfoutput\undefined
\def\rgboo#1{}
\def\postscript#1{\special{" #1}}		
\postscript{
	/bd {bind def} bind def
	/fsd {findfont exch scalefont def} bd
	/sms {setfont moveto show} bd
	/ms {moveto show} bd
	/pdfmark where		
	{pop} {userdict /pdfmark /cleartomark load put} ifelse
	[ /PageMode /UseOutlines		
	/DOCVIEW pdfmark}
\def\bookmark#1#2{\postscript{		
	[ /Dest /MyDest\the\marknumber /View [ /XYZ null null null ] /DEST pdfmark
	[ /Title (#2) /Count #1 /Dest /MyDest\the\marknumber /OUT pdfmark}%
	\advance\marknumber by1}
\def\pdfclink#1#2#3{%
	\hskip-.25em\setbox0=\hbox{#2}%
		\countdp=\dp0 \countwd=\wd0 \countht=\ht0%
		\divide\countdp by65536 \divide\countwd by65536%
			\divide\countht by65536%
		\advance\countdp by1 \advance\countwd by1%
			\advance\countht by1%
		\def\linkdp{\the\countdp} \def\linkwd{\the\countwd}%
			\def\linkht{\the\countht}%
	\postscript{
		[ /Rect [ -1.5 -\linkdp.0 0\linkwd.0 0\linkht.5 ] 
		/Border [ 0 0 0 ]
		/Action << /Subtype /URI /URI (#3) >>
		/Subtype /Link
		/ANN pdfmark}{\rgb{#1}{#2}}}
%
%
\else
\def\rgboo#1{\pdfliteral{#1 rg #1 RG}}
\pdfcatalog{/PageMode /UseOutlines}		
\def\bookmark#1#2{
	\pdfdest num \marknumber xyz
	\pdfoutline goto num \marknumber count #1 {#2}
	\advance\marknumber by1}
\def\pdfklink#1#2{%
	\noindent\pdfstartlink user
		{/Subtype /Link
		/Border [ 0 0 0 ]
		/A << /S /URI /URI (#2) >>}{\rgb{1 0 0}{#1}}%
	\pdfendlink}
\fi

\def\rgbo#1#2{\rgboo{#1}#2\rgboo{0 0 0}}
\def\rgb#1#2{\mark{#1}\rgbo{#1}{#2}\mark{0 0 0}}
\def\pdfklink#1#2{\pdfclink{1 0 0}{#1}{#2}}
\def\pdflink#1{\pdfklink{#1}{#1}}
%
%
\newcount\seccount  
\newcount\subcount  
\newcount\ssscount  
\newcount\clmcount  
\newcount\equcount  
\newcount\refcount  
\newcount\demcount  
\newcount\execount  
\newcount\procount  
\seccount=0
\equcount=0
\clmcount=0
\subcount=0
\refcount=0
\demcount=0
\execount=0
\procount=0
%
\def\proof{\medskip\noindent{\bf Proof.\ }}
\def\proofof(#1){\medskip\noindent{\bf Proof of \csname c#1\endcsname.\ }}
\def\qed{\hfill{\sevenbf QED}\par\medskip}
\def\references{\bigskip\noindent\hbox{\bf References}\medskip
                \ifx\pdflink\undefined\else\bookmark{0}{References}\fi}
\def\addref#1{\global\advance\refcount by 1
              \expandafter\xdef\csname r#1\endcsname{\number\refcount}}

\def\nextremark #1\par{\item{$\circ$} #1}
\def\firstremark #1\par{\bigskip\noindent{\bf Remarks.}
     \smallskip\nextremark #1\par}
\def\abstract#1\par{{\baselineskip=10pt
    \eightpoint\narrower\noindent{\eightbf Abstract.} #1\par}}
%
\def\equtag#1{\global\advance\equcount by 1
              \expandafter\xdef\csname e#1\endcsname{(\number\seccount.\number\equcount)}}
\def\equation(#1){\equtag{#1}\eqno\csname e#1\endcsname}
\def\equ(#1){\hskip-0.03em\csname e#1\endcsname}
%
\def\clmtag#1#2{\global\advance\clmcount by 1
                \expandafter\xdef\csname cn#2\endcsname{\number\seccount.\number\clmcount}
                \expandafter\xdef\csname c#2\endcsname{#1~\number\seccount.\number\clmcount}}
\def\claim #1(#2) #3\par{\clmtag{#1}{#2}
    \vskip.1in\medbreak\noindent
    {\bf \csname c#2\endcsname .\ }{\sl #3}\par
    \ifdim\lastskip<\medskipamount
    \removelastskip\penalty55\medskip\fi}
\def\clm(#1){\csname c#1\endcsname}
\def\clmno(#1){\csname cn#1\endcsname}
%
\def\sectag#1{\global\advance\seccount by 1
              \expandafter\xdef\csname sectionname\endcsname{\number\seccount. #1}
              \equcount=0 \clmcount=0 \subcount=0 \execount=0 \procount=0}
\def\section#1\par{\vskip0pt plus.1\vsize\penalty-40
    \vskip0pt plus -.1\vsize\bigskip\bigskip
    \sectag{#1}
    \message{\sectionname}\leftline{\twelvebf\sectionname}
    \nobreak\smallskip\noindent
    \ifx\pdflink\undefined
    \else
      \bookmark{0}{\sectionname}
    \fi}
%
\def\subtag#1{\global\advance\subcount by 1
              \expandafter\xdef\csname subsectionname\endcsname{\number\seccount.\number\subcount. #1}
              \ssscount=0}
\def\subsection#1\par{\vskip0pt plus.05\vsize\penalty-20
    \vskip0pt plus -.05\vsize\medskip\medskip
    \subtag{#1}
    \message{\subsectionname}\leftline{\tenbf\subsectionname}
    \nobreak\smallskip\noindent
    \ifx\pdflink\undefined
    \else
      \bookmark{0}{... \subsectionname}  
    \fi}
%
\def\ssstag#1{\global\advance\ssscount by 1
              \expandafter\xdef\csname subsubsectionname\endcsname{\number\seccount.\number\subcount.\number\ssscount. #1}}
\def\subsubsection#1\par{\vskip0pt plus.02\vsize\penalty-10
    \vskip0pt plus -.02\vsize\smallskip\smallskip
    \ssstag{#1}
    \message{\subsubsectionname}\leftline{\ninebf\subsubsectionname}
    \nobreak\smallskip\noindent
    \ifx\pdflink\undefined
    \else
      \bookmark{0}{...... \subsubsectionname}  
    \fi}
%
\def\demtag#1#2{\global\advance\demcount by 1
                \expandafter\xdef\csname de#2\endcsname{#1~\number\demcount}}
\def\demo #1(#2) #3\par{
  \demtag{#1}{#2}
  \vskip.1in\medbreak\noindent
  {\bf #1 \number\demcount.\enspace}
  {\rm #3}\par
  \ifdim\lastskip<\medskipamount
  \removelastskip\penalty55\medskip\fi}
\def\dem(#1){\csname de#1\endcsname}
%
\def\exetag#1{\global\advance\execount by 1
              \expandafter\xdef\csname ex#1\endcsname{Exercise~\number\seccount.\number\execount}}
\def\exercise(#1) #2\par{
  \exetag{#1}
  \vskip.1in\medbreak\noindent
  {\bf Exercise \number\execount.}
  {\rm #2}\par
  \ifdim\lastskip<\medskipamount
  \removelastskip\penalty55\medskip\fi}
\def\exe(#1){\csname ex#1\endcsname}
%
\def\protag#1{\global\advance\procount by 1
              \expandafter\xdef\csname pr#1\endcsname{\number\seccount.\number\procount}}
\def\problem(#1) #2\par{
  \ifnum\procount=0
    \parskip=6pt
    \vbox{\bigskip\centerline{\bf Problems \number\seccount}\nobreak\medskip}
  \fi
  \protag{#1}
  \item{\number\procount.} #2}
\def\pro(#1){Problem \csname pr#1\endcsname}
%
%
%
\def\rightheadline{\hfil}
\def\leftheadline{\sevenrm\hfil HANS KOCH\hfil}
\headline={\ifnum\pageno=\firstpage\hfil\else
\ifodd\pageno{{\fiverm\rightheadline}\number\pageno}
\else{\number\pageno\fiverm\leftheadline}\fi\fi}
\footline={\ifnum\pageno=\firstpage\hss\tenrm\folio\hss\else\hss\fi}
\let\ov=\overline
\let\cl=\centerline

\let\eps=\varepsilon
\let\sss=\scriptscriptstyle

\def\AA{{\cal A}}
\def\BB{{\cal B}}
\def\CC{{\cal C}}

\def\FF{{\cal F}}
\def\GG{{\cal G}}

\def\KK{{\cal K}}
\def\LL{{\cal L}}
\def\MM{{\cal M}}
\def\NN{{\cal N}}
\def\OO{{\cal O}}
\def\PP{{\cal P}}

\def\RR{{\cal R}}
\def\SS{{\cal S}}

\def\UU{{\cal U}}
\def\VV{{\cal V}}
\def\WW{{\cal W}}
\def\XX{{\cal X}}
\def\YY{{\cal Y}}

\def\ssA{{\sss A}}
\def\ssB{{\sss B}}

\def\ssF{{\sss F}}
\def\ssG{{\sss G}}

\def\ssP{{\sss P}}

\def\id{{\rm I}}

\def\tr{\mathop{\rm tr}\nolimits}
\def\det{\mathop{\rm det}\nolimits}

\def\diag{\mathop{\rm diag}\nolimits}

\def\Im{\mathop{\rm Im}\nolimits}
%
\newfam\dsfam
\def\mathds #1{{\fam\dsfam\tends #1}}

\font\tends=dsrom10
\font\eightds=dsrom8
\textfont\dsfam=\tends
\scriptfont\dsfam=\eightds
%

\def\integer{{\mathds Z}}
\def\rational{{\mathds Q}}
\def\real{{\mathds R}}
\def\complex{{\mathds C}}

\def\torus{{\Bbb T}}

\def\oo{{\scriptstyle\OO}}
\def\bdot{\hbox{\bf .}}
\def\bcomma{\hbox{\bf ,}}
\def\defeq{\mathrel{\mathop=^{\sss\rm def}}}
\def\half{{1\over 2}}

\def\quarter{{1\over 4}}
\def\thalf{{\textstyle\half}}

\def\twovec#1#2{\left[\matrix{#1\cr#2\cr}\right]}

\def\twomat#1#2#3#4{\left[\matrix{#1&#2\cr#3&#4\cr}\right]}

%

%

%

%

%
\newdimen\savedparindent
\savedparindent=\parindent
\font\tenamsb=msbm10 \font\sevenamsb=msbm7 \font\fiveamsb=msbm5
\newfam\bbfam
\textfont\bbfam=\tenamsb
\scriptfont\bbfam=\sevenamsb
\scriptscriptfont\bbfam=\fiveamsb

\def\AM{{\ninerm AM~}}

\def\RG{{\ninerm RG~}}
\def\PM{\phantom{-}}
\def\rhoF{{\rho_{_{\hskip-0.5pt F}}}}
\def\rhoG{{\rho_{_{\hskip-0.5pt G}}}}

\def\varrhoF{{\varrho_{_{\hskip-0.5pt F}}}}

\def\buc{{\hbox{\magnineeufm c}}}

\def\buC{{\hbox{\teneufm C}}}

\def\buF{{\hbox{\teneufm F}}}

\def\buM{{\hbox{\teneufm M}}}
\def\buN{{\hbox{\teneufm N}}}
\def\buR{{\hbox{\teneufm R}}}

\def\buT{{\hbox{\teneufm T}}}
\def\bus{{\hbox{\magnineeufm s}}}

\font\fourrm=cmr5 at 4pt
\def\srmF{{\hbox{\fiverm F}}}
\def\srmG{{\hbox{\fiverm G}}}
\def\ssrmF{{\hbox{\fourrm F}}}
\def\ssrmG{{\hbox{\fourrm G}}}
\let\ub\underbar   
\def\circle{{\Bbb S}}
\def\col{{:\hskip3pt}}
\def\scol{{;\hskip3pt}}

\def\mod{\mathop{\rm mod}\nolimits}

\def\bfZero{{\bf 0}}
\def\bfOne{{\bf 1}}
\let\idmat\bfOne
\def\rmGL{{\rm GL}}
\def\rmSL{{\rm SL}}

\def\ssA{{\sss A}}
\def\ssB{{\sss B}}

\def\transpose{{\sss\top}}

\def\tinyskip{\hskip.7pt}

\def\oo{{\scriptstyle\OO}}

\def\tfrac#1#2{{\textstyle{#1\over #2}}}
\def\ssspar(#1){{\scriptscriptstyle(}#1{\scriptscriptstyle)}}
\def\sspar(#1){{\scriptstyle(}#1{\scriptstyle)}}
\def\bdot{\hbox{\bf .}}
\def\hdots{\line{\leaders\hbox to 0.5em{\hss .\hss}\hfil}}

\def\sfrac#1#2{\hbox{\raise2.2pt\hbox{$\scriptstyle#1$}\hskip-1.2pt
   {$\scriptstyle/$}\hskip-0.9pt\lower2.2pt\hbox{$\scriptstyle#2$}\hskip1.0pt}}
\def\shalf{\sfrac{1}{2}}
\def\squarter{\sfrac{1}{4}}

\def\stwomat#1#2#3#4{{\eightpoint\left[\matrix{#1&#2\cr#3&#4\cr}\right]}}

\def\today{\ifcase\month\or
January\or February\or March\or April\or May\or June\or
July\or August\or September\or October\or November\or December\fi
\space\number\day, \number\year}
\addref{Harp}
\addref{Sos}
\addref{Sur}
\addref{Swi}
\addref{Hof}
\addref{HPS}
\addref{Fei}
\addref{CouTr}
\addref{Kada}
\addref{TKNdN}
\addref{McK}
\addref{BeSi}
\addref{JoMo}
\addref{PalisMelo}
\addref{ORJS}
\addref{AvSi}
\addref{EKW}
\addref{EckWit}
\addref{LanCC}
\addref{DeFaria}
\addref{Eliass}
\addref{Sullivan}
\addref{WieZa}
\addref{Lastii}
\addref{McMu}
\addref{BES}
\addref{FaKa}
\addref{HKW}
\addref{Furman}
\addref{GJLS}
\addref{RuPi}
\addref{Lyub}
\addref{Jito}
\addref{MOW}
\addref{OsAv}
\addref{Yampol}
\addref{dCLM}
\addref{Dama}
\addref{GoSch}
\addref{AvAC}
\addref{Oseledets}
\addref{GaJo}
\addref{AvJii}
\addref{AvJi}
\addref{AKiKS}
\addref{AKcRG}
\addref{AvKrii}
\addref{AvG}
\addref{Satij}
\addref{KochAP}
\addref{KochAM}
\addref{KKi}
\addref{SatWil}
\addref{KochTrig}
\addref{Files}
\addref{Ada}
\addref{Gnat}
\addref{IEEE}
\addref{MPFR}
\def\leftheadline{\sixrm\hfil H.~Koch\hfil\today}
\def\rightheadline{\sevenrm\hfil Scaling and universality for skew products\hfil}
%
\newfam\bigdsfam
\def\bigmathds #1{{\fam\bigdsfam\bigtwelveds #1}}
\font\bigtwelveds=dsrom12 scaled\magstep1
\font\bigtends=dsrom10 scaled\magstep1
\font\bigeightds=dsrom8 scaled\magstep1
\textfont\bigdsfam=\bigtends
\scriptfont\bigdsfam=\bigeightds
\cl{{\twelvebf Asymptotic scaling and universality}}
\cl{{\twelvebf for skew products with factors in SL(2,$\bigmathds R$)}}
\bigskip

\cl{
Hans Koch
\footnote{$\!^1$}
{\eightpoint\hskip-2.7em
Department of Mathematics, The University of Texas at Austin,
Austin, TX 78712.}}

\bigskip
\abstract
We consider skew-product maps over circle rotations
$x\mapsto x+\alpha\;(\mod\,1)$
with factors that take values in $\rmSL(2,{\scriptstyle\real})$.
This includes maps of almost Mathieu type.
In numerical experiments, with $\alpha$ the inverse golden mean,
Fibonacci iterates of maps from the almost Mathieu family
exhibit asymptotic scaling behavior
that is reminiscent of critical phase transitions.
In a restricted setup that is characterized by a symmetry,
we prove that critical behavior indeed occurs and is universal
in an open neighborhood of the almost Mathieu family.
This behavior is governed by a periodic orbit
of a renormalization transformation.
An extension of this transformation is shown to have
a second periodic orbit as well,
and we present some evidence that this orbit attracts
supercritical almost Mathieu maps.

\section Introduction

We consider the asymptotic behavior of skew products
$$
A^{\ast q}(x)\,\defeq\,
A(x+(q-1)\alpha)\cdots A(x+2\alpha)A(x+\alpha)A(x)\,,
\equation(Aastqx)
$$
as $q\to\infty$ along certain subsequences,
where $A$ is a real analytic function
from the circle $\torus=\real/\integer$ to the group $\rmSL(2,\real)$.
Here, $\alpha$ is a given irrational number
and $x\mapsto x+\alpha$ is considered modulo $1$.
Our main results concern the inverse golden mean
$\alpha=\sqrt{5}/2-1/2$,
but we expect analogous results to hold for arbitrary quadratic irrationals.

The products \equ(Aastqx) arise when iterating a map $G$
on $\torus\times\real^2$ of the form
$$
G(x,y)=\bigl(x+\alpha,A(x)y\bigr)\,,\qquad x\in\torus\,,\quad y\in\real^2\,.
\equation(Gxy)
$$
Such maps will be called skew-product maps.
We will use the notation $G=(\alpha,A)$ and refer to $A$ as the factor of $G$.
In this notation,
the $q$-th iterate of $G$ is $G^q=(q\alpha,A^{\ast q})$,
with $A^{\ast q}$ given by \equ(Aastqx).
We note that the map $G$ is invertible,
with inverse $G^{-1}=\bigl(-\alpha,A(\,\bdot-\alpha)^{-1}\bigr)$.
The $q$-th iterate of $G^{-1}$ will be denoted by $G^{-q}$.

Two dynamical quantities associated with such a skew-product map $G$
are its Lyapunov exponent $L(G)$ and its fibered rotation number $\varrho(G)$.
They are defined by
$$
L(G)=\lim_{q\to\infty}{1\over q}\log\bigl\|A^{\ast q}(x)\bigr\|\,,\qquad
\varrho(G)=\lim_{q\to\infty}{1\over 2\pi q}\arg\GG^q(x,\vartheta)\,,
\equation(LyapRot)
$$
where $\arg(x,\vartheta)=\vartheta$.
Here, $\GG$ denotes a lift of the map
$(x,y)\mapsto\bigl(x+\alpha,\|A(x)y\|^{-1}A(x)y\bigr)$
from $\torus\times\circle$ to $\torus\times\real$,
where $\circle$ denotes the unit circle $\|y\|=1$ in $\real^2$.
Assuming that $A:\torus\to\rmSL(2,\real)$ is continuous and $\alpha$ irrational,
the limit for $\varrho(G)$ does not depend on $x$ or $\vartheta$,
and convergence is uniform.
Furthermore, it is independent modulo $1$ of the choice of the lift $\GG$.
Under the same assumptions, the limit for $L(G)$
exists and is a.e.~constant in $x$.
For proofs of these and related facts we refer to
[\rBeSi,\rJoMo,\rAvSi,\rGoSch].

A skew-product map $G$ is said to be of Schr\"odinger type
if its factor is of the form
$$
A(x)=A((E,s),x)=\twomat{E-\lambda v(x)}{-1}{1}{0}\,,\qquad\lambda=e^s\,.
\equation(SchrA)
$$
A bi-infinite orbit $n\mapsto(x_n,y_n)$ for such a map $G$
has the property that $y_n=\bigl[{u_n\atop u_{n-1}}\bigr]$
for some sequence $n\mapsto u_n$ of real numbers.
If $x_0=0$, then this sequence $u$
is a solution of the equation $H^\alpha_\lambda u=Eu$,
where $H^\alpha_\lambda$ is the Schr\"odinger operator given by the equation
$$
(H^\alpha_\lambda u)_n=u_{n+1}+u_{n-1}+\lambda v(n\alpha)u_n\,,\qquad
n\in\integer\,.
\equation(AMHamiltonian)
$$
The choice of potential $v(x)=2\cos(2\pi(x+\xi))$
defines the family of operators $H^\alpha_\lambda$
that are knows as almost Mathieu ({\ninerm AM)} operators.
They describe the motion of an electron on $\integer^2$
under the influence of a magnetic flux $2\pi\alpha$ per unit cell,
if one restricts to wave functions $\phi(n,m)=u_n e^{-2\pi im\xi}u_n$.
The full Hamiltonian for this system
is known as the Hofstadter Hamiltonian [\rHarp,\rHof].
These operators have been studied extensively over the past $20$ years.
Two reviews can be found in [\rLastii,\rDama].

Some of the most interesting phenomena in physics
arise from the fact that asymptotic quantities
can depend in a nontrivial way on model parameters.
In the \AM family, the main parameters (besides $\alpha$)
are the coupling constant $\lambda$ and the energy $E$.
The asymptotic quantities include the Lyapunov exponent $L$
and the fibered rotation number $\varrho$.
Among the many known properties are the following
[\rGJLS,\rAvJii,\rLastii,\rAvKrii,\rDama].
Here, we suppress the dependence on the parameter $\xi$,
since it is trivial, as was mentioned after \equ(LyapRot).

Assume that $\alpha$ is irrational.
Then the spectrum $\Sigma^\alpha_\lambda$ of the operator $H^\alpha_\lambda$
on $\ell^2(\integer)$
is a Cantor set of measure $2-2\min\{\lambda,1/\lambda\}$.
For all energies in the spectrum,
the Lyapunov exponent of the corresponding \AM map $G$
is given by $L(G)=\max\{0,\log\lambda\}$.
The fibered rotation number $\varrho$
is a continuous decreasing function of the energy $E$,
and it is constant on each spectral gap
\big(a connected component of $\real\setminus\Sigma^\alpha_\lambda$\big).
As was described first in [\rTKNdN,\rJoMo],
this resonance phenomenon has an interesting arithmetic aspect:
each gap can be labeled canonically by an integer $k$, known as the Hall conductance.
On the gap with index $k$, the fibered rotation number
is constant and satisfies $1-2\varrho(G)\equiv k\alpha\;(\mod 1)$.
The left hand side of this congruence
can also be identified with the integrated density of states
[\rBeSi,\rAvSi,\rBES,\rGoSch] for the Hamiltonian $H^\alpha_\lambda$.

\smallskip
Regions where asymptotic quantities depend analytically on
model parameters are also called phases.
By varying the parameters, it is possible to induce phase transitions.
A common phenomenon observed in such transitions is universality:
within a large class of systems,
the type of singularity is independent of the system being considered,
down to precise values of observable quantities.
The theory of critical phenomena aims to explain
situations where the singularities involve power laws.
Power law behavior represents asymptotic scale invariance,
and the quantities that describe such universal scaling
are known as critical exponents.

Similar phenomena have been observed in comparatively simple systems.
Some examples will be mentioned below.
Based on numerical observations and partial results [\rKochAM,\rKKi],
we conjecture that skew-product maps exhibit such universal scaling as well.

To be more specific,
we consider the inverse golden mean $\alpha_\ast=\sqrt{5}/2-1/2$.
Denote by $p_k/q_k$ the $k$-th continued fraction approximant for $\alpha_\ast$.
That is, $p_k$ is the $k$-th Fibonacci number, and $q_k=p_{k+1}$.

\claim Conjecture(Conjecture)
There exists a ``large'' class $\AA$
of real analytic functions $A:\torus\to\rmSL(2,\real)$,
which includes the \AM factors for $\xi=\alpha_\ast/2$,
for which the following holds.
Let $\varrho$ be a rational number in $[0,\sfrac{1}{2}]$.
Then for every real analytic two-parameter family
$\beta\mapsto A(\beta,\bdot)$ of functions in $\AA$
that satisfies a certain transversality condition,
there exists a parameter value $\beta_\ast$
where the map $(\alpha_\ast,A(\beta_\ast,\bdot))$ has fibered rotation number $\varrho$,
as well as three matrices $L,C,M\in\rmGL(2,\real)$,
such that the limits
$$
\eqalign{
B_\ast(\beta,x)
&=\lim_{n\to\infty}L^{-n}A^{\ast p_{\ell n}}
\bigl(\beta_\ast+CM^{-n}\beta\,\bcomma\,\alpha_\ast^{\ell n}x\bigr)L^n\,,\cr
A_\ast(\beta,x)
&=\lim_{n\to\infty}L^{-n}A^{\ast q_{\ell n}}
\bigl(\beta_\ast+CM^{-n}\beta\bcomma\,\alpha_\ast^{\ell n}x\bigr)L^n\,,\cr}
\equation(AEProdConv)
$$
exist for all $x\in\real$ and are independent of the given family.
Here $\ell$ is some positive integer that depends only on $\varrho$.
Furthermore, $L$ is conjugate to some fixed matrix $L_\ell\in\rmGL(2,\real)$,
and $M$ is conjugate to some fixed diagonal matrix $\diag(\mu_1,\mu_2)$.

This conjecture has motivated the work presented in this paper
as well as our earlier work in [\rKochAM,\rKKi,\rKochTrig].
The integers $\ell$ that appear in \equ(AEProdConv) can be obtained
by considering the map on the torus $\torus^2$
given by the matrix $\bigl[{1~1\atop 1~0}\bigr]$.
Every point $(0,\varrho)$ with $\varrho$ rational
lies on a periodic orbit for this map.
The number $\ell=\ell(\varrho)$ is the shortest such period.
For a class $\AA$ that includes the \AM factors,
one finds that $\ell$ must be a multiple of $3$,
due to a symmetry of this family.
For more details we refer to [\rKKi].

\smallskip
We believe that \clm(Conjecture) holds for any irrational $\alpha_\ast$
that has a periodic continued fraction expansion.
(For general quadratic irrationals,
the same should apply after finitely many steps
of the transformation $\buR$ defined below.)
An extended version could include $\alpha$ as a parameter.
Based on renormalization arguments,
we expect a similar scaling in the difference $\alpha-\alpha_\ast$.
But we have not investigated this situation.

\smallskip
An important aspect of \clm(Conjecture) is universality:
near its critical point $\beta_\ast$,
the behavior of a family can be described accurately
in terms of just two parameters.
The limits in \equ(AEProdConv),
as well as the conjugacy class of the matrices $L$ and $M$
are independent of the family.
Universality of this type
plays an important role in the description
of critical phenomena in condensed matter physics,
where it is impossible to know a system precisely.

The points $\beta_\ast$ represent phase transitions for the chosen family.
In the \AM family parametrized by $\beta=(E,s)$,
it is known that the system described by the Hamiltonian $H^\alpha_\lambda$
undergoes a transition from a conducting phase (a.c.~spectrum)
for $\lambda=e^s<1$ to an insulation phase (p.p.~spectrum) for $\lambda=e^s>1$.
For proofs and references we refer to [\rJito].
So in this case, we expect that $s_\ast=0$
at each critical point $\beta_\ast=(E_\ast,s_\ast)$.
Furthermore, our numerical computations suggest that the scaling $M$ is diagonal.

The ``phase portrait'' for $\lambda=1$,
obtained by plotting the spectrum $H^\alpha_1$ as a set-valued function of $\alpha$,
is known as the Hofstadter butterfly [\rHof].
A detailed topological description of the Hofstadter butterfly can be found in [\rOsAv].
One of its striking features, aside from the gap labeling,
is a local self-similarity property:
successive magnifications about certain points
seem to yield an asymptotic limit set [\rRuPi,\rSatij,\rKochAM,\rKKi,\rSatWil].

Cases where the expected phase transitions have been studied
from the point of view of critical phenomena cover the rotation numbers
$\varrho=\sfrac{1}{2},\sfrac{3}{8},\sfrac{2}{6},\sfrac{1}{4},\sfrac{1}{6},\sfrac{1}{8},0$.
The values $\varrho=\sfrac{1}{2}$ and $\varrho=0$
correspond to the energies $E_\ast=\mp 2.5975\ldots$
at the bottom and top of the spectrum, respectively.
The value $\varrho=\squarter$ corresponds to $E_\ast=0$, for symmetry reasons.
These three cases have been considered in [\rKochAM,\rKKi].
In particular, rough numerical computations
indicate that the Hofstadter butterfly is asymptotically invariant under a scaling
about the point $(\alpha_\ast,E_\ast)$,
and that the scaling factor in the energy direction is given by the constant $\mu_1$.

\smallskip
A framework that has been extremely successful
in describing critical phenomena is renormalization.
In the area of dynamical systems,
this includes period-doubling cascades for interval maps
[\rFei,\rCouTr,\rEckWit,\rEliass,\rSullivan,\rMcMu,\rLyub]
or area-preserving maps [\rEKW,\rdCLM,\rGaJo],
critical circle mappings [\rORJS,\rLanCC,\rDeFaria,\rYampol],
and the breakup of invariant tori in area-preserving maps
[\rKada,\rMcK,\rAKcRG,\rKochAP], to name just a few.

In the problem at hand,
we expect renormalization to work as follows.
The functions $B_n$ and $A_n$ whose limits
are being considered in \equ(AEProdConv)
are the factors associated with two skew-product maps
$F_n$ and $G_n$ that depend on a parameter $\beta$.
In the renormalization framework, the sequence of pairs $P_n=(F_n,G_n)$
lie on an orbit of a transformation $\buR$ that acts on a space of pairs.
The accumulation property \equ(AEProdConv)
describes convergence (modulo re-parametrization) $P_n\to P_\ast$
to a fixed point $P_\ast$ of $\buR^\ell$.
The fixed point $P_\ast$ and the observed accumulation rates are universal,
due to the fact that they reflect properties of the transformation $\buR$.
In particular, we expect $\buR^\ell$ to be hyperbolic at $P_\ast$,
with a two-dimensional local unstable manifold.
This manifold is given by the family of pairs
$\beta\mapsto{\tt P}_\ast(\beta)$ whose factors
are limit functions $B_\ast$ and $A_\ast$ in \equ(AEProdConv).
The transversality condition mentioned in \clm(Conjecture)
requires that the given family be transversal to the stable manifold of $\buR^\ell$.
And $\beta_\ast$ is the value of the parameter $\beta$
where a given family intersects the stable manifold.

\smallskip
Our goal here is to verify this renormalization picture
in a setup that is restricted but includes most of the essential aspects.
After finding an appropriate transformation $\buR$,
the first step in any renormalization group ({\ninerm RG}) analysis
is to prove the existence of a small invariant set, such as a periodic orbit.
For the type of skew-product maps considered here,
a fixed point of $\buR^3$ was obtained in [\rKochAP]
for $\varrho\in\{\sfrac{1}{2},0\}$.
A theorem concerning the existence of a fixed point of $\buR^6$
associated with $\varrho=\squarter$ was announced in [\rKKi].
A proof of this theorem will be given in Section 4.

Constructing a fixed point $P_\ast$ for $\buR^\ell$ is a local analysis
(near an approximate fixed point).
By contrast, proving that the \AM family and others are attracted
to the unstable manifold of $\buR^\ell$ at $P_\ast$ is a global analysis and much harder.
A simplified version of this problem was considered in [\rKochTrig],
in a situation where the Hofstadter Hamiltonian reduces
to skew-product maps with factors that take values
in the circle $\real\cup\{\infty\}$.
Here we prove convergence \equ(AEProdConv) with $\rmSL(2,\real)$ factors,
including the \AM factors, but only in a restricted one-parameter setup.

\smallskip
Before describing our main results,
we would like to mention a peculiarity of these skew-product maps.
High accuracy computations suggest [\rKKi]
that the eigenvalue $\mu_1$ associated with $\varrho\in\{\sfrac{1}{2},0\}$
is a zero of the polynomial $\PP_6(z)=z^4-196z^3-58z^2-4z+1$.
And the eigenvalue $\mu_1$ associated with $\varrho=\squarter$
is a zero of $\PP_3(z)=z^4-30z^3-24z^2-10z-1$.
In both cases, the product of the two real roots of $\PP_\ell$ is $(-\alpha)^{-\ell}$.
Furthermore, the value of $\mu_2$ appears to be $\alpha^{-\ell}$.
In fact, computations that were carried out in the context of the present paper
suggest that the eigenvalues of $D\buR^\ell(P_\ast)$ can all be written down in algebraic form.
It is unusual that universal constants associated with critical phenomena
are algebraically related to basic system parameters,
such as the flux parameter $\alpha$ here.
Known exceptions are statistical mechanics models in $2$ dimensions,
where the large scale asymptotic is governed by a conformal symmetry.
The class of skew-product maps
that includes the \AM maps seems to be governed by symmetries as well,
but it is not clear how the symmetries of the Hofstadter model [\rWieZa,\rFaKa,\rHKW]
generate the algebraic eigenvalues that are observed here.

\section Main results

As is common in the renormalization of maps that include a circle rotation,
we first generalize the notion of periodicity by considering commuting pairs of maps.
Consider the map $F=(1,\idmat)$ on $\real\times\real^2$, defined by $F(x,y)=(x+1,y)$.
A skew-product map $G=(\alpha,A)$ with a factor $A:\real\to\rmSL(2,\real)$
represents a map on on the cylinder $\torus\times\real^2$
if and only if $G$ commutes with $F$.
A more general skew-product map $F=(1,B)$ on $\real\times\real^2$
can be viewed as defining a cylinder $\torus_\ssF\times\real^2$
embedded in $\real\times\real^2$, by identifying points on the orbit of $F$.
If $G$ commutes with $F$,
then  $G$ defines a map on this cylinder $\torus_\ssF\times\real^2$.

Consider now pairs $(F,G)$
of maps $F=(1,B)$ and $G=(\alpha,A)$ on $\real\times\real^2$ that commute.
Here, $\alpha$ can be an arbitrary irrational number between $0$ and $1$.
Then the renormalized pair is defined by the equation
$$
\buR((F,G))=\bigl(\check F,\check G\bigr)\,,\qquad
\check F=\Lambda^{-1}G\Lambda\,,\quad
\check G=\Lambda^{-1}FG^{-c}\Lambda\,,
\equation(RGDef)
$$
where $c$ is the integer part of $\alpha^{-1}$,
and where $\Lambda(x,y)=\bigl(\alpha x,L_1y\bigr)$.
Here, $L_1$ is a suitable nonsingular $2\times 2$ matrix
that can chosen to depend on the pair $(F,G)$.
By construction, the first component of $\check F$ is again $1$,
while the first component of $\check G$ is $\check\alpha=\alpha^{-1}-c$.
We note that $\alpha\mapsto\check\alpha$ is the Gauss map
that appears in the continued fraction expansion of $\alpha$.

In what follows, $\alpha$ is assumed to be the inverse golden mean.
Its continued fraction expansion is $\alpha=1/(1+1/(1+\ldots))$,
so $\alpha$ is a fixed point of the Gauss map, and $c=1$ in the equation \equ(RGDef).
As mentioned earlier, we expect to find a period of $\buR$ that is a multiple of $3$.
This leads us to consider
orbits of the third iterate of $\buR$, which is of the form
$$
\buR^3(P)=\bigl(\Lambda_3^{-1}G^2F^{-1}\Lambda_3\,\bcomma\,
\Lambda_3 FG^{-1}FG^{-2}\Lambda_3\bigr)\,,\qquad
P=(F,G)\,,
\equation(RGiii)
$$
with $\Lambda_3(x,y)=\bigl(\alpha^3 x,L_3y\bigr)$
for some suitable nonsingular $2\times 2$ matrix $L_3$.

What plays an important role in our analysis are symmetry properties.
A $2\times 2$ matrix $\Sigma$ will be called a reflection,
if $\Sigma^2=\idmat$ and $\det(\Sigma)=-1$.
An invertible map $H$ on $\real\times\real^2$ is said to be reversible
with respect to $\Sigma$, if
$$
H^{-1}=\SS H\SS\,,\qquad\SS(x,y)=(-x,\Sigma y)\,.
\equation(Greversible)
$$
For a skew-product map $H=(\gamma,C)$,
reversibility with respect to $\Sigma$ is equivalent to the property
$$
C_\circ(x)^{-1}=\Sigma C_\circ(-x)\Sigma\,,\qquad
C_\circ(x)\defeq C\bigl(x-\tfrac{\gamma}{2}\bigr)\,.
\equation(AoReversible)
$$
The matrix-valued function $C_\circ$ defined by \equ(AoReversible)
will be referred to as the symmetric factor of $H$,
even if $H$ is not reversible.
A pair $P=(F,G)$ will be called reversible
if both $F$ and $G$ are reversible with respect to the same reflection $\Sigma$.
Since the matrix $\Sigma$ depends on a choice of coordinates,
we will specify it only when necessary.

The following result was announced in [\rKochAM].
A proof will be given in Subsection 4.3.

\claim Theorem(ExistenceSix)
Let $\alpha$ be the inverse golden mean.
Then $\buR^6$ has a reversible fixed point $P_\star=(F_\star,G_\star)$
with $F_\star=(1,B_\star)$ and $G_\star=(\alpha,A_\star)$ commuting.
The factors $B_\star$ and $A_\star$ are non-constant
entire functions with values in $\rmSL(2,\real)$.
The scaling $L_6$ at $P_\star$ has real eigenvalues $\VV$ and $\VV^{-1}$
whose sum is $2\alpha^{-3}$, up to an error less that $10^{-429}$,

To be more precise, the scaling matrix $L_6$ mentioned in this theorem
is the product of the matrix $L_3(P_\star)$
appearing in the transformation $P_\star\mapsto P=\buR^3(P_\star)$,
and the matrix $L_3(P)$
appearing in the transformation $P\mapsto P_\star=\buR^3(P)$.
The exact form of $L_6$ depends on the chosen coordinates.

As a by-product of our (computer-assisted) proof of this theorem,
we have accurate bounds on the various quantities involved,
as well as other numerical data.
These data include approximate values for the two expanding
eigenvalue $\mu_1$ and $\mu_2$ of $D\buR^6(P_\star)$.

As will be described in Section 4,
the \AM family for $\xi=\sfrac{\alpha}{2}$ is reversible,
due to the fact that $x\mapsto E-\lambda\cos\bigl(2\pi x)$ is an even function.
By choosing the $y$-scaling $L_3$ appropriately,
reversibility (for a fixed $\Sigma$) is preserved under renormalization.
So we expect \clm(Conjecture) to hold within a class
of reversible pairs.
The fixed point $P_\star$ described in \clm(ExistenceSix)
is associated with the fibered rotation number $\varrho=\squarter$.
In the \AM family, this corresponds to the energy $E_\ast=0$.

Our main goal in this paper is to prove \clm(Conjecture) in a simplified setting
where the analysis can be restricted to one-parameter families.
This lead us to consider maps that are anti-reversible.
To be more precise, define $-(\gamma,C)=(\gamma,-C)$.
Using the same notation as in \equ(Greversible),
we say that $H$ is anti-reversible with respect to $\Sigma$, if $H^{-1}=-\SS H\SS$.
A pair $(F,G)$ is said to be anti-reversible,
if $F$ is reversible and $G$ anti-reversible
with respect to the same reflection $\Sigma$

If we choose $\xi=\sfrac{\alpha}{2}-\sfrac{1}{4}$,
then the \AM map $G$ is anti-reversible, but only for $E=0$,
due to the fact that
$x\mapsto E-\lambda\sin\bigl(2\pi x)$ is an odd function precisely when $E=0$.
So the idea is to restrict our analysis to anti-reversible pairs.
By choosing the $y$-scaling $L_3$ appropriately,
anti-reversibility (for a fixed $\Sigma$) is preserved under renormalization.
So we expect \clm(Conjecture) to hold for one-parameter families
in this restricted class,
except that the \RG transformation has only a single expanding direction.

Based on \clm(ExistenceSix), we expect to find a fundamental period $6$
in this case. Somewhat unexpectedly, we find a period $3$.

\claim Theorem(MapRGFix)
Let $\alpha$ be the inverse golden mean.
Then $\buR^3$ has an anti-reversible fixed point $P_\star=(F_\star,G_\star)$
with $F_\star=(1,B_\star)$ and $G_\star=(\alpha,A_\star)$ commuting.
The factors $B_\ast$ and $A_\ast$ are non-constant
entire functions with values in $\rmSL(2,\real)$.
The scaling $L_3$ at $P_\ast$ is an orthogonal reflection in $\real^2$
about some line (that depends on the choice of coordinates).
An extension of $\buR_3$ to pairs that need not commute is hyperbolic,
with a single expanding direction with eigenvalue $\mu_2\ge\alpha^{-3}$.

A proof of this theorem will be given in Subsection 5.1.

\demo Remark(NegStep)
In the anti-reversible case,
our \RG transformations $\buR^3$ and $\buR_3$
include an extra step $(B,A)\mapsto(-B,-A)$.
We will ignore this step here, e.g.~by identifying pairs of factors up to a sign.

For the pair $P_\ast$ and the eigenvalue $\mu_2$
described in \clm(MapRGFix) we also have the following.
Here, and in the remaining part of this paper,
$\alpha$ always denotes the inverse golden mean, unless specified otherwise.

\claim Theorem(ProdConv)
Consider the \AM factors \equ(SchrA) with $\xi=\sfrac{\alpha}{2}-\sfrac{1}{4}$
and energy $E=0$.
Denote by  $p_n$ the $n$-th Fibonacci number and let $q_n=p_{n+1}$.
Then there exists an open disk $D\subset\complex$ centered at the origin,
such that the limits
$$
\eqalign{
B_\ast(s,x)
&=\lim_{n\to\infty}L_3^{-1}A^{\ast p_{3n}}\bigl(\mu_2^{-3n}s,\alpha^{3n}x\bigr)L_3\,,\cr
A_\ast(s,x)
&=\lim_{n\to\infty}L_3^{-1}A^{\ast q_{3n}}\bigl(\mu_2^{-3n}s,\alpha^{3n}x\bigr)L_3\,,\cr}
\equation(AEProdConv)
$$
exist for all $s\in D$ and all $x\in\complex$.
Here, $L_3$ is some orthogonal reflection in the plane.
The functions $(s,x)\mapsto A_\ast(s,x)$
and $(s,x)\mapsto B_\ast(s,x)$ are analytic on $D\times\complex$,
and the convergence in \equ(AEProdConv) is uniform
on compact subsets of this domain.
The family of pairs $s\mapsto{\tt P}_\ast(s)$
associated with the limit factors \equ(AEProdConv)
is a parametrization of the local unstable manifold of $\buR_3$ at $P_\ast$.
Furthermore, the same holds for any real-analytic family ${\tt P}$
in some open neighborhood (in a suitable topology) of the \AM family,
after an initial affine re-parametrization $s\mapsto s_\ast+cs$
with $c\ne 0$.

A proof of this theorem will be given in Subsection 5.2.
To be more precise,
our proof of Theorems \clmno(ExistenceSix), \clmno(MapRGFix), and \clmno(ProdConv)
is computer-assisted.
This means that some estimates have been verified
(rigorously) with the aid of a computer.
The main steps and ideas are described in Section 8.
For details we refer to the source code of our programs [\rFiles].

The main part of our analysis is carried out in a space $\FF_\rho$ of pairs of maps
$F=(1,B)$ and $G=(\alpha,A)$ whose symmetric factors $A_\circ$ and $B_\circ$
are analytic in a bounded domain $|x|<\rhoF$ and $|x|<\rhoG$, respectively.
The ``suitable topology'' mentioned in \clm(ProdConv) only compares
factors on this bounded domain; so in particular, the factors need not be periodic.
Entire analyticity of $A_\ast$ and $B_\ast$ is obtained a-posteriori
from the fact that the  transformation $\buR^3$ is analyticity-improving.
(And it is not hard to see that these factors are of finite exponential type.)

\smallskip
To be more specific, we consider the fixed point problem for $\buR_3$
instead of $\buR^3$, where
$$
\buR_3(P)=\bigl(\Lambda_3^{-1}GF^{-1}G\Lambda_3\,\bcomma\,
\Lambda_3 G^{-1}FG^{-1}FG^{-1}\Lambda_3\bigr)\,.
\equation(PaliRGiii)
$$
This makes no difference for commuting pairs.
But for non-commuting pairs, which need to be included in our analysis,
the transformation $\buR^3$ does not in general preserve (anti)reversibility,
while $\buR_3$ does.
After constructing a fixed point $(F_\ast,G_\ast)$ for the transformation $\buR_3$,
we can use \equ(AEProdConv) to conclude that $F_\ast$ and $G_\ast$ commute.
Our extension of $\buR_3$ to nearly-commuting pairs
also includes a ``commutator correction''
which makes this transformation contracting in the direction
of non-commuting perturbations.

Hyperbolicity of $\buR_3$ is proved via estimates
on the derivative $D\buR_3$ on some cylinder $C_1'$ centered at $P_\ast$.
In order to prove \clm(ProdConv),
we show that some \RG iterate of the anti-reversible \AM family
defines a curve that is properly aligned with a cylinder $C_0'\subset C_1'$.
This constitutes the global part of our analysis.
What remains is again a purely local problem.

One of the claims in \clm(ProdConv) is that the parameter value $s_\ast$
for which the \AM pair is attracted to the fixed point $P_\ast$ is zero.
This is specific to the \AM family and has to be proved separately.
Similarly, our guess that $\mu_2=\alpha^{-3}$
is based on special properties of the \AM family.
A proof is again outside the scope of renormalization.

In our proof that $s_\ast=0$, we use the fact that
the Lyapunov exponent of the \AM map $G$ for a spectral energy
is $L(G)=\max\{0,\log\lambda\}$.
The general idea is that $q\mapsto G^q$ tends to infinity if $L(G)$ is positive.
An argument along these lines shows that $s_\ast\le 0$.
Proving that $s_\ast\ge 0$ turns out to be significantly harder.

A useful tool in our proof of \clm(ProdConv)
is a Lyapunov exponent for pairs $P=(F,G)$.
This exponent $L(P)$ is defined in such a way that
it agrees with $L(G)$, if $F=(1,\idmat)$ and $G=(\alpha,A)$,
with $\alpha$ the inverse golden mean.
It also has the property that
$$
L\bigl(\buR(P)\bigr)=\alpha^{-1}L(P)\,.
\equation(RGLyap)
$$
This shows e.g.~that $L(P_\ast)=0$.
The equation \equ(RGLyap) also suggests that $\mu_2=\alpha^{-3}$.
Unfortunately, we can only prove that $\mu_2\ge\alpha^{-3}$.

The problem of proving $\mu_2\le\alpha^{-3}$
is related to the question of whether $L$ takes a positive value
on the local unstable manifold of $\buR_3$ at $P_\ast$.
Our pursuit of this question has led to some
interesting observations that we shall now describe.

\smallskip
Given that $L(P)$ is an asymptotic quantity,
it is necessary to consider the unstable manifold $\WW^u$ globally,
at least on the side where we expect $L(P)$ to be positive.
Based on numerical experiments,
our conjecture is that $\WW^u$ gets attracted
to a ``supercritical'' fixed point $P_\diamond$.
In fact, all \AM pairs with $\lambda>1$ and $E=0$ appear to be get attracted
to this fixed point.

To be more specific, we have to describe an extension of $\buR_3$
to pairs of skew-product maps whose factors need not have determinant $1$.
Let $H=(\gamma,C)$.
If $\det(C)$ is the constant function $x\mapsto 1$,
then the inverse $H^{-1}$ of $H$ agrees with the quasi-inverse
$$
H^\dagger=\bigl(-\alpha,C^\dagger(\,\bdot-\alpha)\bigr)\,,
\quad{\rm where}\quad
C^\dagger=\stwomat{d}{-b}{-c}{a}
\quad{\rm if}\quad
C=\stwomat{a}{b}{c}{d}\,.
\equation(GpApDef)
$$
Thus, we can extend the domain of $\buR_3$
by replacing the inverse maps in our definition \equ(PaliRGiii) by their quasi-inverses.
But we assume that the determinants are nonnegative.
In our proof of Theorems \clmno(ExistenceSix) and \clmno(MapRGFix),
we use such an extension of $\buR_3$ for pairs whose factors
have determinants close to $x\mapsto 1$.
The extended transformation $\buR_3$ includes (as its last step) a normalization
that divides each factors by the square root of its determinant.
So any fixed point of $\buR_3$ or $\buR_3^2$
has factors that take values in $\rmSL(2,\real)$.

In what follows, we allow factors
that (are nonzero but) can have arbitrary nonnegative constant determinants.
But our \RG transformation now includes a normalization step
that divides each factor by its norm.
This is useful in cases where the norms would otherwise
tend to infinity under iteration of $\buR_3$.
For such an extension we find the following.

\claim Theorem(SuperCritFix)
There exists entire functions $b_\diamond$ and $a_\diamond$ of order $1$,
with $x\mapsto b_\diamond(x-\sfrac{1}{2})$ even
and $x\mapsto a_\diamond(x-\sfrac{\alpha}{2})$ odd,
such that the pair $P_\diamond=((1,B_\diamond),(\alpha,A_\diamond))$ with
$$
B_\diamond(x)=b_\diamond(x)\stwomat{0}{0}{0}{1}\,,\qquad
A_\diamond(x)=a_\diamond(x)\stwomat{1}{0}{0}{0}\,,
\equation(SuperCritFix)
$$
is a fixed point of $\buR_3$ with $L_3=\bigl[{0~1\atop 1~0}\bigr]$.
The zeros of $b_\diamond$ are all simple and define
a non-periodic bi-infinite sequence of real numbers
whose gaps take exactly three distinct values:
$\shalf$, $\alpha^{-1}$, and $\alpha^{-1}+\shalf$.
The zeros of $a_\diamond$ have an analogous property,
except that the gaps only take two distinct values:
$\shalf$ and $\alpha^{-1}-\shalf$.

Numerically, we find that $P_\diamond$ attracts supercritical \AM pairs,
as well as pairs ${\tt P}(s)$ with $s>0$ on the unstable manifold of $P_\ast$.
Our computations covered several values of $\lambda$ between $1+2^{-32}$ and $2$,
both for $\xi=\sfrac{\alpha}{2}-\sfrac{1}{4}$ and $\xi=\sfrac{\alpha}{2}$.
To be more precise, the $y$-scaling has to include a rotation;
otherwise the limit can be a rotated version of $P_\diamond$.
We would expect similar behavior
for other energies in the spectrum of $H^\alpha_\lambda$,
as well as for other quadratic irrationals $\alpha$
that have a periodic continued fraction expansion.

For the inverse golden mean,
it should be possible to prove that the (anti)reversible \AM pair
with $\lambda$ sufficiently large is attracted to $P_\diamond$ under iteration of $\buR_3$,
but such an analysis would go beyond the scope of this paper.
A strong-coupling fixed point for an approximate
renormalization scheme has been constructed in [\rMOW].

What we will prove here is the following.

\claim Theorem(AMZeros)
Let $P_0$ be an anti-reversible \AM pair with coupling constant $\lambda>1$.
Consider the set of accumulation points
of the sequence $n\mapsto\buR^{3n}(P_0)$ in the space $\FF_\rho$ mentioned earlier.
This set $K_\ast$ is compact and invariant under $\buR_3$.
Let $P=((1,B),(\alpha,A))$ be any pair in $K_\ast$.
Then $A$ and $B$ extend to entire functions,
with $b_\circ=\tr(B_\circ)$ even and $a_\circ=\tr(A_\circ)$ odd.
Neither $B$ nor $A$ are constant, and $B$ is symmetric.
Furthermore, $B(x)=0$ wherever $b_\diamond(x)=0$,
and $A(x)=0$ wherever $a_\diamond(x)=0$.

A proof of Theorems \clmno(SuperCritFix) and \clmno(AMZeros)
will be given in Section 7.

\smallskip
We have not investigated the asymptotic behavior of subcritical pairs,
like the \AM pairs for $\lambda<1$.
Results on almost-reducibility [\rAvJi]
suggest that such pairs converge to some $\buR$-invariant set
that consists of pairs whose factors are constant.
The action of $\buR$ on pairs with constant factors is trivial.
In particular, it is easy to find periodic orbits
for any rational fibered rotation number $\varrho$.
Whether or not \AM pairs with non-small positive coupling constant $\lambda<1$
and rational fibered rotation number $\varrho$
converge to such a ``subcritical'' fixed point
(of $\buR^\ell$ for some $\ell$) is a global question
and not easy to answer.

\section The RG transformation for anti-reversible pairs

The main goal in this section is to properly formulate
the fixed point problem considered in \clm(MapRGFix).
Our \RG analysis of anti-reversible pairs
will be continued in Section 5,
after having covered the reversible case in Section 4.

\subsection Some basic facts and identities

Let $P=(F,G)$ be a pair of skew-product maps
$F=(1,B)$ and $G=(\alpha,A)$ whose factors $B$ and $A$
take values in $\rmGL(2,\real)$ and have positive determinants.
Then the renormalized pair $\tilde P=\buR_3(P)$ is given by
$$
\tilde P=\bigl(\tilde F,\tilde G\bigr)\,,\qquad
\tilde F=\Lambda_3^{-1}\hat F\Lambda_3\,,\qquad
\tilde F=\Lambda_3^{-1}\hat F\Lambda_3\,,
\equation(BasicRGThree)
$$
where
$$
\hat F=GF^\dagger G\,,\qquad
\hat G=G^\dagger FG^\dagger FG^\dagger\,.
\equation(hatFhatG)
$$
Here, $F^\dagger$ and $G^\dagger$ denote the quasi-inverses
of $F$ and $G$, respectively, as defined in \equ(GpApDef).
The first component of $\hat F$ is $2\alpha-1=\alpha^3$.
So after scaling by $\alpha^3$,
the first component of $\tilde F$ is again $1$.
Similarly, the first component of $\hat G$ is $2-3\alpha=\alpha^4$.
So after scaling by $\alpha^3$,
the first component of $\tilde G$ is again $\alpha$.
The symmetric factor $\hat B$ of $\hat F$ is given by
$$
\textstyle
\hat B_\circ(x)=A_\circ\bigl({\alpha-1\over 2}+x\bigr)B_\circ(x)^\dagger
A_\circ\bigl({1-\alpha\over 2}+x\bigr)\,,
\equation(MatGFiG)
$$
and for the symmetric factor $\hat A$ of $\hat G$ we obtain
$$
\textstyle
\hat A_\circ(x)=A_\circ\bigl((1-\alpha)+x\bigr)^\dagger
B_\circ\bigl({1-\alpha\over 2}+x\bigr)A_\circ(x)^\dagger
B_\circ\bigl({\alpha-1\over 2}+x\bigr)
A_\circ\bigl((\alpha-1)+x\bigr)^\dagger\,.
\equation(MatGiFGiFGi)
$$
The symmetric factors associated with $\tilde P=\buR_3(P)$
are now obtained via scaling:
$$
\tilde B_\circ(x)=L_3^{-1}\hat B_\circ\bigl(\alpha^3 x\bigr)L_3\,,\qquad
\tilde A_\circ(x)=L_3^{-1}\hat A_\circ\bigl(\alpha^3 x\bigr)L_3\,.
\equation(TildeSymmHatSymm)
$$
Notice that such a relation is obvious for the regular factors.
But it holds for the symmetric factors as well, as a short computation shows.
Our choice for the $y$-scaling matrices $L_3$ will be described
in Subsection 3.2.

Next, let us consider some consequences of (anti)reversibility.
To this end, and for reference later on, define
$$
J=\stwomat{0}{1}{-1}{0}\,,\qquad
S=\stwomat{1}{0}{0}{-1}\,,\qquad
M=2^{-1/2}\stwomat{1}{1}{1}{-1}\,.
\equation(JSMDef)
$$
With the exception of Section 4, (anti)reversibility in this paper is defined
with respect to the reflection $\Sigma=iJ$.
Notice that conjugacy by $iJ$ keeps real matrices real.

Assume now that $F=(1,B)$ is reversible and $G=(\alpha,A)$ anti-reversible,
both with respect to $\Sigma=iJ$.
A short computation shows that this condition is equivalent to
$$
B_\circ(x)^\transpose=B_\circ(-x)\,,\qquad
A_\circ(x)^\transpose=-A_\circ(-x)\,.
\equation(RevBoAntiAo)
$$
Here, $C^\transpose$ denotes the transpose of a matrix $C$.

What makes skew-products over irrational rotations difficult to deal with
is that products $A^{\ast q}(x)$ with large $q$ can vary vastly in size, as a function of $x$.
If the Lyapunov exponent $L=L(G)$ is positive,
then $A^{\ast q}(x)$ grows asymptotically like $e^{qL}$ for typical values of $x$.
Particularly large factors can obtained via the identity
$$
(-1)^m A_\circ^{\ast 2m}(iy)
=U(iy)^\ast U(iy)\,,
\equation(asEvenPowerAoiy)
$$
where
$U(x)=A_\circ((m-\sfrac{1}{2})\alpha+x)
\cdots A_\circ(\sfrac{\alpha}{2}+x)$,
and where $U^\ast=\ov{U}^\transpose$ denotes the adjoint of $U$.
So in particular, $(-1)^m A_\circ^{\ast 2m}(iy)$ is a positive matrix for $y\in\real$.
This fact will be used in several of our proofs.

On the other hand, products with many factors can be of order $1$ in size.
Consider the case where $\det(A)=1$.
Using that $A_\circ(-x)=-J^{-1}A_\circ(x)^{-1}J$, we have
$$
(-1)^m A^{\ast(2m+1)}_\circ(x)
=V(x)A_\circ(x)J^{-1}V(-x)^{-1}J\,,
\equation(asOddPowerAo)
$$
where $V(x)=A_\circ((m-1)\alpha+x)\cdots A_\circ(\alpha+x)$.
If $A_\circ(0)=-J$, then this implies that
$$
(-1)^m A_\circ^{\ast(2m+1)}(0)=-J\,.
\equation(asOddPowerAoZero)
$$
This applies e.g.~to the \AM map with $\xi=\sfrac{\alpha}{2}-\sfrac{1}{4}$ and $E=0$.
And it is independent of the value of $\lambda$.
So for $\lambda>1$,
sub-products that appear in $A^q$ that are of the form \equ(asOddPowerAoZero)
are much smaller than sub-products of the form \equ(asEvenPowerAoiy) for $y=0$.
This is the mechanism that produces the zeros described in \clm(AMZeros).

In the remaining part of this paper,
we consider the \AM maps with respect to the basis
defined by the column vectors of the matrix $M$ given in \equ(JSMDef).
In this representation, the symmetric factor
of the anti-reversible \AM map is given by
$$
A_\circ=\twomat{t_\circ}{t_\circ+1}{t_\circ-1}{t_\circ}\,,\qquad
t_\circ(x)=\lambda\sin(2\pi x)\,.
\equation(SinAMo)
$$

\subsection Scaling and normalization

As mentioned earlier, we choose $L_3$ to be a reflection matrix.
In the coordinates considered here,
$$
L_3=L(\vartheta)\defeq\twomat
{\cos(\vartheta+\pi/4)}
{-\sin(\vartheta+\pi/4)}
{-\sin(\vartheta+\pi/4)}
{-\cos(\vartheta+\pi/4)}\,.
\equation(Ltheta)
$$
Notice that $L(\vartheta)=L(0)e^{-\vartheta J}$,
where $e^{-\vartheta J}$ is a rotation by $\vartheta$.
Notice also that $L_3^2=\idmat$.
So $L_3$ drops out in a fixed point equation for $\buR_3^2$.
But if $P=(F,G)$ is a fixed point of $\buR_3^2$,
then a conjugacy by any rotation yields another fixed point.

The goal is to get uniqueness
by choosing $\vartheta=\vartheta(P)$
in such a way that $\tilde P=\buR_3(P)$ satisfies
a suitable normalization condition.
With a $y$-scaling $L_3$ of the form \equ(Ltheta),
the symmetric factor of $\tilde F=\Lambda_3^{-1}\hat F\Lambda_3$
is given by
$$
\tilde B_\circ(0)
=e^{\vartheta J}\stwomat{a}{u}{v}{d}e^{-\vartheta J}\,,\qquad
\stwomat{a}{u}{v}{d}\defeq -L(0)\hat B_\circ(0)L(0)\,,
\equation(AntiRevEqualize)
$$
where $\hat B_\circ$ is as described in \equ(MatGFiG).
The negative sign on the right hand side of this equation
is due to the step $(B,A)\mapsto(-B,-A)$ mentioned in \dem(NegStep).
As a normalization condition,
we impose that the two entries on the main diagonal of $\tilde B_\circ(0)$ agree.
A straightforward computation shows that
this determines $\buc=\cos(2\vartheta)$ and $\bus=\sin(2\vartheta)$
as follows:
$$
\bus={a-d\over q}\,,\qquad\buc=-{u+v\over q}\,,\qquad
q=\sqrt{(u+v)^2+(a-d)^2}\,.
\equation(busFour)
$$
As tedious as such computations may be,
explicit expressions like \equ(busFour) are needed
in a computer-assisted proof that is by nature highly constructive.
In order to get explicit expressions for the derivative
$D\buR_3(P)\dot P$, it is convenient
to consider a family of pairs $P$ that depend differentiably on a parameter.
Then the quantities $a,u,v,d$ defined by \equ(AntiRevEqualize)
depend differentiably on the parameter as well.
Using the ``dot notation'' for derivatives with respect to the parameter,
the derivatives of $\bus$ and $\buc$ are given by
$$
\dot\bus={\buc\over q}
\bigl[\buc\bigl(\dot a-\dot d\bigr)+\bus(\dot u+\dot v)\bigr]\,,\qquad
\dot\buc=-{\bus\over q}
\bigl[\buc\bigl(\dot a-\dot d\bigr)+\bus(\dot u+\dot v)\bigr]\,.
\equation(dotbus)
$$

At this point we have defined the ``basic'' version of our \RG
transformation $\buR_3$ for skew-product pairs $P=(F,G)$.
We are not assuming that $F=(1,B)$ and $G=(\alpha,A)$ commute,
nor that $A$ and $B$ have determinant $1$.
But the transformation does not behave as desired for non-commuting pairs
or for factors that have determinants $\ne 1$.
Denote this basic version by $\RR_3$.
Our extended version of $\buR_3$ is defined as
$$
\buR_3=\buN\circ\RR_3\circ\buC\,,
\equation(FullMapRG)
$$
where $\buC$ is a ``commutator correction''
that will be defined later,
and where $\buN$ performs a re-normalization of determinants.

To normalize determinants, we simply choose
$$
\buN\bigl((\gamma,C)\bigr)=\bigl(\gamma,\NN(C)\bigr)\,,\qquad
\NN(C)=[\det(C)]^{-1/2}C\,.
\equation(DivSqrtDet)
$$
If the determinant of $C$ is close to $1$,
then \equ(DivSqrtDet) is well-defined and $\NN(C)$ has determinant $1$.
We note that, if $H=(\gamma,C)$ is (anti)reversible,
then $\det(C)$ is an even function, so $\buN(H)$ is still (anti)reversible.
For a pair $P=(F,G)$ we define $\buN$ component-wise.

For estimates of derivatives $D\buR_3(P)\dot P$,
we use that the derivative of $\NN$ at
$C=\bigl[{a~u\atop v~d}\bigr]$ is given by
$$
D\NN(C)\dot C=
\det(C)^{-1/2}\dot C
-\thalf\det(C)^{-3/2}\bigl[a\dot d+d\dot a-u\dot v-v\dot u\,\bigr]C\,.
\equation(DerNN)
$$

\subsection Commutators

The linearization of the basic transformation $\RR_3$
at the fixed point $P_\ast$ can have non-contracting directions
that are associated with non-commuting perturbations of $P_\ast$.
Formally, one can see that $D\RR_3(P_\ast)$ must have an eigenvalue $-1$.
And numerically, another eigenvalue is the number $\VV=8.3524100320\ldots$
that appears in \clm(ExistenceSix).
The goal is to eliminate these two eigenvalues.
This will be done in the next section.

First we need some generalities.
Consider the commutator $\Theta=FG(GF)^{-1}$ for a pair $P=(F,G)$.
A straightforward computation shows that
the commutator for the renormalized pair $\tilde P=\bigl(\tilde F,\tilde G\bigr)$
is given by
$$
\tilde\Theta=(G\Lambda_3)^{-1}\Theta^{-1}(G\Lambda_3)\,.
\equation(tildeThetaAgain)
$$
If we write $\Theta=(0,C)$ and $\tilde\Theta=\bigl(0,\tilde C)$,
then
$$
\tilde C(x)=L(\vartheta)^{-1}
A\bigl(\alpha^3x\bigr)^{-1}C\bigl(\alpha^3x+\alpha\bigr)^{-1}
A\bigl(\alpha^3x\bigr)L(\vartheta)\,.
\equation(artildeThetamatOne)
$$
Consider the change of variables $x={1+\alpha\over 2}+z$
and define
$$
\CC(P,z)=\textstyle C\bigl({1+\alpha\over 2}+z\bigr)\,,\qquad
\AA(P,z)=A_0\bigl(\thalf+\alpha^3z\big)L(\vartheta_\ssP)\,.
\equation(CCPz)
$$
Then the equation \equ(artildeThetamatOne) becomes
$$
\CC\bigl(\tilde P,z\bigr)
=\AA(P,z)^{-1}\CC\bigl(P,\alpha^3 z\bigr)^{-1}\AA(P,z)\,.
\equation(artildeThetamatThree)
$$
{}From this equation one can see that the eigenvalues
of $\Theta\mapsto\tilde\Theta$ at $\Theta=\id$
are determined by the behavior of $\CC(P,z)$ near $z=0$.

Thus, consider $\CC(P)=\CC(P,0)=C\bigl({1+\alpha\over 2}\bigr)$.
An explicit computation shows that
$$
\textstyle
\CC(P)
=XY^{-1}\,,\qquad
X=B_0\bigl({\alpha\over 2}\bigr)
A_0\bigl(-{1\over 2}\bigr)\,,\qquad
Y=A_0\bigl({1\over 2}\bigr)B_0\bigl(-{\alpha\over 2}\bigr)\,.
\equation(CommXYi)
$$
Assume now that $G$ is anti-reversible with respect to $\Sigma=iJ$.
Then $JXJ=Y^{-1}$. So $X$, $Y$, and $XY^{-1}$ are of the form
$$
X=\stwomat{a}{b}{c}{d}\,,\qquad
Y=-\stwomat{a}{c}{b}{d}\,,\qquad
XY^{-1}=\stwomat{-1+b(b-c)}{-a(b-c)}{d(b-c)}{-1-c(b-c)}\,,
\equation(AntiRevXYMat)
$$
with $ad-bc=1$.
If $XY^{-1}$ is the identity matrix, then $X=Y=J$.
So in our applications, the matrix elements $a$ and $d$ are close to zero.

\subsection Commutator corrections

The commutator correction map $\buC$
is the first step in our \RG transformation \equ(FullMapRG).
The goal is for $P'=\buC(P)$ to satisfy $\CC(P')=\idmat$.
We define $\buC$ as a composition of three maps.
To simplify notation, each of these maps will be denoted by $P\mapsto P'$.

\medskip\noindent
\ub{Step 1}. Here we replace the symmetric factor $A_\circ$ of $G$ by
$$
A_\circ'=RA_\circ R\,,\qquad
R=\stwomat{\rho}{r}{r}{\rho}\,,\qquad
\rho^2=1+r^2\,,
\equation(arFaccAcicrc)
$$
while keeping $B_\circ'=B_\circ$.
The goal is to choose $r$ in such a way that $\tr(X')=0$.
Notice that $R$ is reversible, in the sense that $J^{-1}RJ=R^{-1}$.
Since $G$ is anti-reversible,
this guarantees that the map $G'$ is anti-reversible as well.
Write
$$
A_\circ\bigl(-\thalf\bigr)
=\twomat{t_\ssA+s_\ssA}{u_\ssA}{v_\ssA}{t_\ssA-s_\ssA}\,,\qquad
B_\circ\bigl(\tfrac{\alpha}{2}\bigr)
=\twomat{t_\ssB+s_\ssB}{u_\ssB}{v_\ssB}{t_\ssB-s_\ssB}\,,
\equation(arFaccAoBo)
$$
and define
$$
\eqalign{
\eps&=\tr(X_\circ)\,,\cr
\tau&=4t_\ssA t_\ssB+(u_\ssA+v_\ssA)(u_\ssB+v_\ssB)\,,\cr
\sigma&=(u_\ssA+v_\ssA)2t_\ssB+2t_\ssA(u_\ssB+v_\ssB)\,.\cr}
\equation(Farepstausigma)
$$
A tedious but trivial computation shows that $\tr(X')=0$, if we choose
$$
r={-\eps\over
\sqrt{
\half\bigl(\sigma^2-2\eps\tau\bigr)
+\half\sqrt{\bigl(\sigma^2-2\eps\tau\bigr)^2-4\bigl(\tau^2-\sigma^2\bigr)\eps^2}}}\,.
\equation(rFarEquFive)
$$
And for the derivative with respect to a parameter, we obtain
$$
\dot r=-{\dot\eps+r(\rho\dot\sigma+r\dot\tau)\over\varphi}\,,\qquad
\varphi=\sigma\rho+r\bigl(2\tau+r\sigma\rho^{-1}\bigr)\,.
\equation(dotr)
$$

\medskip\noindent
\ub{Step 2}. Assume now that $\tr(X)=0$.
The second correction $P\mapsto P'$ is defined via a transformation
$$
A_\circ'=\KK A_\circ\KK\,,\qquad
B_\circ'=\KK^{-1} B_\circ\KK^{-1}\,,\qquad
\KK=\twomat{\kappa^{1/2}}{0}{0}{\kappa^{-1/2}}\,,
\equation(KKAKKKKiBKKKi)
$$
and the goal is to have
$$
X'=\stwomat{s}{w}{-w}{-s}\qquad{\rm if}\quad
X=\stwomat{s}{b}{c}{-s}\,.
\equation(XbulletOne)
$$
Recall that $X\approx J$ in our applications,
so that $s\approx 0$, $b\approx 1$, and $c\approx-1$.
Clearly $X'=\KK^{-1}X\KK$ is of the desired form, if we choose
$$
\kappa=\sqrt{-b/c}\,.
\equation(cckappa)
$$
Then $w=\sqrt{-bc}$.
The derivative with respect to a parameter is trivial,
so we will not give it here.

\medskip\noindent
\ub{Step 3}. Assume now that $X=\bigl[{\PM s~\PM w\atop-w~-s}\bigr]$.
The third correction $P\mapsto P'$ is of the form
$$
A_\circ'=RA_\circ R\,,\qquad B_\circ'=R^{-1}B_\circ R^{-1}\,,
\qquad R=\stwomat{\rho}{r}{r}{\rho}\,,
\equation(RARRiRRi)
$$
with $\rho^2=1+r^2$.
The goal is to determine $r$ in such a way that $X'=R^{-1}XR$ is equal to $J$.
An explicit computation show that this is achieved with
$$
r={-s\over\sqrt{2\bigl(w^2-s^2\bigr)+2w\sqrt{w^2-s^2}}}\,.
\equation(tarrFive)
$$
For the derivative with respect to a parameter, we find that
$$
\dot r=-{\thalf\dot s+r(\dot w\rho+\dot sr)\over\psi}\,,\qquad
\psi=w\rho+r\bigl(2s+w\rho^{-1}r\bigr)\,.
\equation(tardotrFour)
$$

\subsection The fixed point problem

Consider now the transformation $\buR_3$ defined by \equ(FullMapRG).
Our first goal is to prove that $\buR_3$ has a fixed point $P_\ast$
that has potentially the properties described in \clm(MapRGFix).
As is common in many computer-assisted proofs,
we associate with the given transformation $\buR_3$
a quasi-Newton map $\buM$ that we hope to be
a contraction near some approximate fixed point $\bar P$.
Picking an approximate inverse $\id-M$
of $\id-D\buR_3\bigl(\bar P\bigr)$, we define
$$
\buM(p)=\buR_3\bigl(\bar P+(\id-M)p\bigr)-\bar P+Mp\,.
\equation(ContrRGThree)
$$
Here, the sum of map-pairs is defined component-wise,
and $c_1(\gamma,C_1)+c_2(\gamma,C_2)$ is defined as $(\gamma,c_1C_1+c_2C_2)$.
Notice that, if $p$ is a fixed point of $\buM$,
then $P=\bar P+(\id-M)p$ is a fixed point of $\buR_3$.

The following function spaces have already been used in [\rKochAP].
Given $\rho>0$, denote by $\GG_\rho$ the space of all
real analytic functions $g$ on $(-\rho,\rho)$ that have a finite norm
$$
\|g\|_\rho=\sum_{n=0}^\infty|g_n|\rho^n\,,\qquad
g(x)=\sum_{n=0}^\infty g_n x^n\,.
\equation(AANorm)
$$
Notice that every function $g\in\GG_\rho$
extends analytically to the complex disk $|x|<\rho$.
Furthermore, $\GG_\rho$ is a Banach algebra
under the pointwise product of functions.

The space of matrix functions
$$
C_\circ=\twomat{t_\circ+s_\circ}{u_\circ}{v_\circ}{t_\circ-s_\circ}\,,
\equation(RepTUVS)
$$
with $t_\circ$, $u_\circ$, $v_\circ$, and $s_\circ$
belonging to $\GG_\rho$ will be denoted by $\GG_\rho^4$.
The norm of $C_\circ\in\GG_\rho^4$ is defined as
$\|C_\circ\|_\rho=\|t_\circ\|_\rho+\|u_\circ\|_\rho+\|v_\circ\|_\rho+\|s_\circ\|_\rho$.

Given a pair $\rho=(\rho_\srmF,\rho_\srmG)$ of positive real numbers,
we define $\FF_\rho$ to be the vector space of all pairs
$\PP=(B_\circ,A_\circ)$ in $\GG_{\rho_\ssrmF}^4\times\GG_{\rho_\ssrmG}^4$,
equipped with the norm
$\|\PP\|_\rho=\|B_\circ\|_{\rho_\ssrmF}+\|A_\circ\|_{\rho_\ssrmG}$.
The subspace of pairs $\PP\in\FF_\rho$ that satisfy the (anti)reversibility
conditions \equ(RevBoAntiAo) will be denoted by $\FF_\rho^{\tinyskip r}$.

For simplicity, and when no confusion can arise,
we will identify a skew-product map $H=(\gamma,C)$
with its symmetric factor $C_\circ$.
Referring to the representation \equ(RepTUVS),
we note that $H$ is reversible with respect to $iJ$, if and only if
the functions $t_\circ$ and $s_\circ$ are even, while $v_\circ(-x)=u_\circ(x)$.
Or $C_\circ$ is anti-reversible,
precisely if $t_\circ$ and $s_\circ$ are odd, while $v_\circ(x)=-u_\circ(-x)$.

In our applications, we always choose $\varrho_\ssF\le\varrho_\ssG$.
Under these conditions, \equ(MatGFiG) and \equ(MatGiFGiFGi)
show that $\buR_3$ is well-defined on $\FF_\rho$ if
$$
\thalf<\rho_\srmG\le
\rho_\srmF<\alpha^{-3}\rho_\srmG-\thalf\alpha^{-1}\,.
\equation(RRThreeDomainCond)
$$
To be more precise, the conditions needed in the normalization step $\buN$
and for the commutator correction $\buC$
(all of which represent ad-hoc choices)
also require some mild nondegeneracy properties.
We note that the domain conditions for the transformation $\buR$
are more restrictive than the conditions \equ(RRThreeDomainCond) for $\buR_3$.
But both are satisfied with comfortable margins
in the case $\rho_\ssF=2$ and $\rho_\ssG=\sfrac{11}{8}$ considered below.

For reference later on, we note that the transformation $\buR_3$ is compact,
due to the analyticity-improving property of $\buR_3$.
To be more precise the transformation
$(B_\circ,A_\circ)\mapsto\bigl(\tilde B_\circ,\tilde A_\circ\bigl)$
defined by the equations \equ(MatGFiG), \equ(MatGiFGiFGi), and \equ(TildeSymmHatSymm)
maps bounded sets in $\FF_\rho$ to bounded sets in $\FF_{\rho'}$,
for some choice of $\rho_\ssF'>\rho_\ssF$ and $\rho_\ssG'>\rho_\ssG$.
And the inclusion map from $\FF_{\rho'}$ into $\FF_\rho$ is compact.

\claim Lemma(AntiRevContr)
Let $\rho=(2,\sfrac{11}{8})$.
Then there exist a pair $\bar P$ in $\FF_\rho^{\tinyskip r}$,
a bounded linear operator $M$ on $\FF_\rho^{\tinyskip r}$,
and positive constants $\eps,K,\delta$ satisfying
$\eps+K\delta<\delta$, such that
the transformation $\buM$ defined by \equ(ContrRGThree)
is analytic in $B_\delta$ and satisfies
$$
\|\buM(0)\|_\rho\le\eps\,,\qquad
\|D\buM(p)\|_\rho\le K\,,\qquad p\in B_\delta\,,
\equation(ContrBounds)
$$
where $B_\delta$ denotes the open ball of radius $\delta$
in $\FF_\rho^{\tinyskip r}$, centered at the origin.
Every pair $p\in B_\delta$ has the following properties.
The matrix components of $P=\bar P+(\id-M)p$ are non-constant
and satisfy the bound $\bigl\|P-\bar P\bigr\|_\rho<10^{-450}$.
Furthermore, the angle $\vartheta=\vartheta(P)$
satisfies $\sin(2\vartheta)=-0.01760801\ldots$

Our proof of this lemma is computer-assisted
and will be described in Section 8.
These estimates will be used in Subsection 5.1 to give a proof of \clm(MapRGFix).

\demo Remark(thetaVal)
The angle $\vartheta$ mentioned in \clm(AntiRevContr)
depends on the choice of coordinates.
So it seems to say something about the \AM model, but it is not clear what.
In this context, we note that
the change of coordinates $M$ which yields \AM factors of the form \equ(SinAMo)
achieves nothing useful in the case $\xi=\sfrac{\alpha}{2}-\squarter$
considered here.
It was chosen since it diagonalizes the $y$-scaling in the
reversible case $\xi=\sfrac{\alpha}{2}$.

\section The RG transformation for reversible pairs

The goal here is to reduce the proof of \clm(ExistenceSix)
to technical estimates similar to those in \clm(AntiRevContr).
An extra step is necessary to prove that the pair $P_\star$ commutes.

\subsection Scaling and normalization

We work in a basis where the \AM factor $A_\circ$ for $\xi=\sfrac{\alpha}{2}$
and $E=0$ takes the form \equ(SinAMo), with the sine replaced by a cosine.
The corresponding \AM map $G$ is reversible with respect to $\Sigma=S$,
with $S$ as defined in \equ(JSMDef).
So throughout this section,
we restrict to pairs that are reversible with respect to $\Sigma=S$.
Referring to \equ(RepTUVS), reversibility of $H=(\gamma,C)$
is equivalent to the functions $t_\circ,u_\circ,v_\circ$ being even and $s_\circ$ odd.

The matrix $L_3$ that enters the definition
$\Lambda_3(x,y)=\bigl(\alpha^3x,L_3 y\bigr)$ of the scaling
used for $\buR^3$ and $\buR_3$ is taken to be of the form
$$
L_3=Se^{\sigma_3 S}=\stwomat{e^{\sigma_3}}{0}{0}{-e^{-\sigma_3}}\,,
\equation(RevLiii)
$$
with $\sigma_3=\sigma_3(P)$ depending on the pair $P$ being renormalized.
Notice that $L_3$ commutes with $S$,
so conjugacy by $\Lambda_3$ preserves reversibility.

Instead of $\buR^6$, we first consider the transformation
$$
\buR_6=\buR_2^2\circ\buC\,,\qquad
\buR_3=\buN\circ\RR_3\,,
\equation(RevRGSix)
$$
where $\RR_3$ is the ``basic'' \RG transformation $P\mapsto\tilde P$
defined by the equations \equ(BasicRGThree) and \equ(hatFhatG).
The transformation $\buN$ re-normalizes determinants,
as described after \equ(DivSqrtDet).
The transformation $\buC$ is a commutator correction
that will be described below.

With a $y$-scaling $L_3$ of the form \equ(RevLiii),
the symmetric factor of $\tilde F=\Lambda_3^{-1}\hat F\Lambda_3$
is given by
$$
\tilde B_\circ(0)=\Lambda_3^{-1}\hat B_\circ(0)\Lambda_3
=\twomat{a}{-e^{-2\sigma_3}u}{-e^{2\sigma_3}v}{d}\,,\qquad
\twomat{a}{u}{v}{d}\defeq\hat B_\circ(0)\,,
\equation(RevEqualize)
$$
where $\hat B_\circ$ is as described in \equ(MatGFiG).
We determine $\sigma_3=\sigma_3(P)$ is such a way
that the off-diagonal elements of $\tilde B_\circ(0)$ are equal in modulus.
In other words, $e^{-2\sigma_3}|u|=e^{2\sigma_3}|v|$.
Unless $uv=0$, which does not occur in the cases considered,
this trivially determines the scaling exponent $\sigma_3(P)$.

\subsection Commutator correction

Unlike in the anti-reversible case,
the largest eigenvalue of $\buR_3^2(P_\star)$
in the non-commuting direction appears to be $1$.
Our goal here is to eliminate this eigenvalue.
One reason is that an eigenvalue $1$ makes a quasi-Newton map ill-defined.
Another reason is that the correction will be needed
to prove that the pair $P_\star$ is in fact commuting.

The commutator for $P=(F,G)$ at $x={1+\alpha\over 2}$
is again given by the equation \equ(CommXYi).
By reversibility, we have $SXS=Y^{-1}$.
So $X$, $Y$, and $XY^{-1}$ are of the form
$$
X=\stwomat{a}{b}{c}{d}\,,\qquad
Y=\stwomat{d}{b}{c}{a}\,,\qquad
XY^{-1}=\idmat+(a-d)\stwomat{a}{-b}{c}{-d}\,,
\equation(RevXYMat)
$$
with $ad-bc=1$.
In particular $\tr\bigl(XY^{-1}\bigr)=1+(a-d)^2$.
So reversibility implies that the trace of the commutator
does not change to first order.
This motivate the following.

Consider a commutator correction $\buC:P\mapsto P'$ of the form
$$
A_\circ'=RA_\circ R\,,\qquad B_\circ'=R^{-1}B_0R^{-1}\,,\qquad
R=\stwomat{\rho}{r}{r}{\rho}\,,
\equation(ccAR)
$$
with $\rho^2=1+r^2$.
Then $\buC(P')=X'{Y'}^{-1}$, with $X'=R^{-1}XR$ and $Y'=RYR^{-1}$.
Ideally, we can find $r$ in such away that $X'=Y'$,
or equivalently, that
$$
Y=R^{-2}XR^2\,.
\equation(ccY)
$$
As it tuns out, this can be achieved not only to first order,
but exactly, by choosing
$$
r={q\over 2}\left({2\over(1-q^2)^{1/2}+1-q^2}\right)^{1/2}\,,
\qquad
q=-{a-d\over c-b}\,.
\equation(reqFive)
$$
This completes the definition \equ(RevRGSix) of the transformation $\buR_6$.
For estimates of the derivative $D\buR_6(P)\dot P$ we use that
$$
\dot r={\dot q\over 2}\cdot{\rho\over 1-q^2}\,,\qquad
\dot q={\bigl(\dot a-\dot d\,\bigr)-q\bigl(\dot c-\dot b\bigr)\over c-b}\,.
\equation(dotrdotq)
$$

\subsection Proof of \clm(ExistenceSix)

In this subsection we give a proof of \clm(ExistenceSix)
based on estimates that have been verified with the aid of a computer.
We start by solving the fixed point equation for $\buR_6$.
To this end, we use again a quasi-Newton map
of the type \equ(ContrRGThree), namely
$$
\buM(p)=\buR_6\bigl(\bar P+(\id-M)p\bigr)-\bar P+Mp\,,
\equation(ContrRGSix)
$$
where $\bar P$ is an approximate fixed point of $\buR_6$,
and where $\id-M$ is an approximation for the inverse of
$\id-D\buR_3\bigl(\bar P\bigr)$.
The relevant functions spaces are the spaces
$\GG_\rho$ and $\FF_\rho$ defined in Subsection 2.5.
But $\FF_\rho^{\tinyskip r}$ now denotes the subspace of $\FF_\rho$
of pairs that are reversible with respect to $S$.

\claim Lemma(RevContr)
Let $\rho=(3,2)$.
Then there exist a pair $\bar P$ in $\FF_\rho^{\tinyskip r}$,
a bounded linear operator $M$ on $\FF_\rho^{\tinyskip r}$,
and positive constants $\eps,K,\delta$ satisfying
$\eps+K\delta<\delta$, such that
the transformation $\buM$ defined by \equ(ContrRGSix)
is analytic in $B_\delta$ and satisfies
$$
\|\buM(0)\|_\rho\le\eps\,,\qquad
\|D\buM(p)\|_\rho\le K\,,\qquad p\in B_\delta\,,
\equation(ContrBounds)
$$
where $B_\delta$ denotes the open ball of radius $\delta$
in $\FF_\rho^{\tinyskip r}$, centered at the origin.
Every pair $p\in B_\delta$ has the following properties.
The matrix components of $P=\bar P+(\id-M)p$ are non-constant
and satisfy the bound $\bigl\|P-\bar P\bigr\|_\rho<10^{-439}$.
Furthermore, the six-step scaling factor $\VV=e^{\sigma_6(P)}$
satisfies the bound described in \clm(ExistenceSix).

Our proof of this lemma is computer-assisted
and will be described in Section 8.

By the contraction mapping theorem,
\clm(RevContr) guarantees the existence of a fixed point
$p_\star\in B_\delta$ for $\buM$
and thus a fixed point $P_\star=\bar P+(\id-M)p_\star$ for $\buR_6$.
The symmetric factors for $F_\ast$ and $G_\ast$
are analytic in the disks $|x|<\rho_\ssF$ and $|x|<\rho_\ssG$, respectively.
A trivial computation, using the expressions \equ(MatGFiG)
and \equ(MatGiFGiFGi) for the symmetric factors
of $\hat F=GF^\dagger G$ and $\hat G=G^\dagger FG^\dagger FG^\dagger$, respectively,
shows that radii of the domains of analyticity increase
with each iteration of $\buR_3$ by a factor larger than $1$.
The factor approaches $\alpha^{-3}$ as the number of iterations increases.
This shows that $B_\star$ and $A_\star$ extend to entire functions.

\smallskip
What remains to be proved is that the components $F_\star$ and $G_\star$
of the pair $P_\star$ commute.
To this end, consider the commutator factor $\CC(P,z)$ defined by \equ(CCPz).
It admits a representation \equ(artildeThetamatThree), with
$$
\AA(P,z)=A_0\bigl(\thalf+\alpha^3z\big)Se^{\sigma_3(P)S}\,.
\equation(AAPzZero)
$$
Let $P_3=\buR_3(P)$ and $P_6=\buR_3(P_3)$.
Applying the identity \equ(artildeThetamatThree) twice, we obtain
$$
\CC(P_6,z)=\AA_2(P,z)^{-1}
\CC\bigl(P,\alpha^6 z\bigr)\AA_2(P,z)\,,
\equation(AAPzSix)
$$
where
$$
\AA_2(P,z)=\AA\bigl(P,\alpha^3z\bigr)\AA(P_3,z)\,.
\equation(AAiiDef)
$$
Let $\CC(P)=\CC(P,0)$.

Consider now the pair $P=\buC(P_\star)$.
Then $\CC(P)$ is the identity matrix.
This follows from our definition of the commutator correction $\buC$.
Given that $P_\star$ is a fixed point of $\buR_6=\buR_3^2\circ\buC$
and thus $P_6=P_\star$,
we see from \equ(AAPzSix) that $\CC(P_\ast)$ is the identity matrix as well.
This implies in particular that $P=P_\star$,
so that $P_\star$ is a fixed point of $\buR_3^2$.

In the case $P=P_\star$, the equation \equ(AAPzSix)
is a linear fixed point equation for the function $z\mapsto\CC(P_\star,z)$.
We already know that $\CC(P_\star,z)$ is the identity matrix for $z=0$.
Whether or not the same holds for $z\ne 0$
depends on the eigenvalues of the matrix $\AA_2(P_\star)=\AA_2(P_\star,0)$.

\claim Lemma(AAEigen)
The eigenvalues of $\AA_2(P_\star)$ are $\nu=2.8900536382\ldots$
and $\nu^{-1}$.

Our proof of this lemma is computer-assisted,
as will be described in Section 8.
It also verifies that the origin $z=0$ belongs to the domain
of analyticity of the function that appear in \equ(AAPzSix).
But this could easily be checked by hand as well.

We note that $\nu$ appears to satisfy the equation $\nu+\nu^{-1}=2\alpha^{-1}$.
If this is the case, and if the scaling factor $\VV$ in \clm(ExistenceSix)
satisfies $\VV+\VV^{-1}=2\alpha^{-3}$, then $\nu^2=\VV$.

In some open neighborhood of the origin in $\complex$,
we have either $\buC(P_\star,z)=\idmat$ for all $z$,
or else
$$
\CC(P_\ast,z)=\idmat+z^n\bigl[\CC_n+\oo(1)\bigr]\,,
\equation(PzIdmatPert)
$$
for some nonzero matrix $\CC_n$ and some integer $n\ge 1$.
Substituting this expression for $\CC(P_\ast,z)$
into \equ(AAPzSix) yields the identity
$$
\CC_n=\alpha^{6n}\AA_2(P_\star)^{-1}\CC_n\AA_2(P_\star)\,.
\equation(PzPertEqu)
$$
The eigenvalues of $\CC_n\mapsto\alpha^{6n}\AA_2(P_\star)^{-1}\CC_n\AA_2(P_\star)$
are $\alpha^{6n}$ and $\alpha^{6n}\nu^{\pm 2}$.
They are all less than $1$,
so the equation \equ(PzPertEqu) cannot have a solution $\CC_n\ne 0$.
This shows that the commutator of $F_\star$ and $G_\star$
is constant and equal to the identity
in some open neighborhood of $x={1+\alpha\over 2}$.
Given that $B_\star$ and $A_\star$ are entire analytic,
this implies that $F_\star$ and $G_\star$ commute.

At this point, the proof of \clm(ExistenceSix) is reduced
to the task of verifying the bounds in Lemmas \clmno(RevContr) and \clmno(AAEigen).

\section Hyperbolicity

Here we consider again the anti-reversible case and the \AM maps \equ(SinAMo).

\subsection Proof of \clm(MapRGFix)

Our goal here is to prove \clm(MapRGFix),
with the exception of the inequality $\mu_2\ge\alpha^{-3}$,
based on estimates that can be (and have been) verified
with the aid of a computer.
The inequality $\mu_2\ge\alpha^{-3}$ will be proved in Section 6.

By the contraction mapping theorem,
\clm(AntiRevContr) guarantees the existence of a fixed point
$p_\ast\in B_\delta$ for $\buM$
and thus a fixed point $P_\ast=\bar P+(\id-M)p_\ast$ for $\buR_3$.
For the same reasons as in the reversible case,
the factors $B_\star$ and $A_\star$ associated with $P_\star$
extend to entire functions.

In order to prove hyperbolicity and related properties,
we consider the transformation $\buT$ defined by
$$
\buT(p)=\LL^{-1}\bigl[\buR_3^2\bigl(P_\ast+\LL p\bigr)-P_\ast\,\bigr]\,,
\equation(buFDef)
$$
where $\LL$ is a suitable linear isomorphism of $\FF_\rho^{\tinyskip r}$.
Clearly $p_\ast=0$ is a fixed point of $\buT$.
We expect the derivative $D\buT(0)$ to have an eigenvalue $\alpha^{-6}$
and no other spectrum outside the open unit disk.
Thus, we consider a decomposition $\FF_\rho^{\tinyskip r}=\UU\oplus\WW$,
where $\UU$ is a convenient one-dimensional subspace of $\FF_\rho^{\tinyskip r}$.
We will refer to $\UU$ and $\WW$ as the vertical and horizontal subspaces, respectively.
Now the isomorphism $\LL$ is chosen in such a way
that the expected expanding direction of $D\buT(0)$ is roughly vertical.
Writing an element $q\in\FF_\rho^{\tinyskip r}$ as $q=\bigl[{u\atop w}\bigr]$,
with $u\in\UU$ and $w\in\WW$, we obtain a representation
$$
D\buT(p)q=\twomat{M_{uu}(p)}{M_{uw}(p)}{M_{wu}(p)}{M_{ww}(p)}
\twovec{u}{w}\,,\qquad q=\twovec{u}{w}\,.
\equation(DbuF)
$$
By choosing $\LL$ properly, the operators $M_{uw}(p):\WW\to\UU$
and $M_{wu}(p):\UU\to\WW$ can be made small for all $p$ near $p_\ast=0$.
And $M_{uu}(p)$ should be close to $\alpha^{-6}\simeq 18$.
In order to simplify notation,
we identify $\UU$ with $\real$ by choosing a unit vector $u_0\in\UU$
and identifying the vector $tu_0$ with the coefficient $t$.

Specific estimates are obtained in terms of an enclosure
$$
N_{uu}^{-}\le M_{uu}(p)\le N_{uu}^{+}
\equation(DbuFMuuBound)
$$
and upper bounds
$$
\|M_{uw}(p)\|\le N_{uw}\,,\qquad
\|M_{wu}(p)\|\le N_{wu}\,,\qquad
\|M_{ww}(p)\|\le N_{ww}\,,
\equation(DbuFBounds)
$$
that hold for all pairs $p$ in a suitable cylinder $C_1$.
Here, and in what follows, $\|\bdot\|$ denotes the norm in $\FF_\rho$.
To be more precise, we determine two cylinders $C_0$ and $C_1$,
such that
$$
C_0\subset C_1\,,\qquad C_j=[-h_j,h_j]\times\{w\in\WW: \|w\|<r\}\,,
\equation(Cylinders)
$$
with $0<r<h_0<h_1$.
Notice that both cylinders are centered at $p_\ast=0$.
The goal is to show that $\buT$ maps $C_0$ into $C_1$
and $C_1\setminus C_0$ into the complement of $C_0$.
To this end, it suffices to prove that
$$
\eqalign{
N_{uu}^{+}h_0+N_{uw}r&<h_1\,,\qquad
N_{wu}h_0+N_{ww}r<r\,,\cr
N_{uu}^{-}h_0-N_{uw}r&>h_0\,.\cr}
\equation(DbufCjBounds)
$$
Here, we have used that $\buT(p+q)-\buT(p)=\int_0^1 D\buT(p+sq)q\,ds$
~whenever $p$ and $p+q$ both belong to $C_1$.

\claim Lemma(CylinderBounds)
There exists a linear isomorphism $\LL$ of $\FF_\rho^{\tinyskip r}$,
as well as positive real numbers
$r<h_0<h_1,N_{uu}^{-}<N_{uu}^{+},N_{uw},N_{wu},N_{ww}$
that satisfy \equ(DbufCjBounds),
such that the derivative of the transformation $\buT$
defined by \equ(buFDef)
satisfies the bounds \equ(DbuFMuuBound) and \equ(DbuFBounds)
for every $p\in C_1$.
So $\buT$ maps $C_0$ into $C_1$
and $C_1\setminus C_0$ to the complement of $C_0$,
with room to spare for taking interiors and/or closures.
Let
$$
a_{\pm}=N_{uu}^{\pm}\pm N_{uw}\,,\quad
b=N_{wu}+N_{ww}\,,\quad
c=N_{uu}^{-}-N_{uw}-N_{wu}-N_{ww}\,.
\equation(abcDef)
$$
Then $a_{+}<19$, $b<\quarter$, and $c>17$.

Our proof of this lemma is computer-assisted,
as will be described in Section 8.

One of the consequences of the ``uniform hyperbolicity''
described in \clm(CylinderBounds) is the following.

\claim Corollary(CylUniq)
Let $p\in C_0$.
Then either $\buT^n(p)$ belongs to $C_1\setminus C_0$ for some $n>0$,
or else $\buT^n(p)\to 0$ as $n\to\infty$.

\proof
Let $p\in C_0$.
Since the orbit of $p$ can exit $C_0$ only via the set $C_1\setminus C_0$,
it suffices to consider the case where $p_n=\buR^n(p)$
belongs to $C_0$ for all $n\ge 0$.

Write $p_n=\bigl[{u_n\atop w_n}\bigr]$ with $u_n\in\UU$ and $w_n\in\WW$.
{}From \equ(DbuFMuuBound) and \equ(DbuFBounds) we see that
$$
|u_{n+1}|\ge N_{uu}^{-}|u_n|-N^{uw}\|w_n\|\,,\qquad
\|w_{n+1}\|\le N_{wu}|u_n|+N_{ww}\|w_n\|\,,
\equation(SMOne)
$$
for all $n\ge 0$.
Assume for contradiction that $|u_m|-\|w_m\|>0$ for some $m\ge 0$.
Then
$$
|u_{m+1}|-\|w_{m+1}\|
\ge N_{uu}^{-}|u_m|-N_{uw}\|w_m\|-N_{wu}|u_m|-N_{ww}\|w_m\|>c\|w_m\|\,,
\equation(SMTwo)
$$
with $c>0$ as defined in \equ(abcDef).
So we have $|u_n|-\|w_n\|>0$ for all $n\ge m$.
Combining this with the first inequality in \equ(SMOne),
we find that
$$
|u_{n+1}|>a|u_n|\,,
\equation(SMThree)
$$
for all $n\ge m$, with $a=a_{-}$ as defined in \equ(abcDef).
Given that $a>1$, this leads to a contradiction.
So we must have $|u_n|\le\|w_n\|$ for all $n\ge 0$.
By the second inequality in \equ(SMOne), this implies that
$$
\|w_{n+1}\|\le N_{wu}|u_n|+N_{ww}\|w_n\|\le b\|w_n\|\,,
\equation(SMFour)
$$
for all $n\ge 0$, with $b$ as defined in \equ(abcDef).
Given that $b<1$, we find that $w_n\to 0$ as $n\to\infty$.
But $|u_n|\le\|w_n\|$ for all $n\ge 0$, so $u_n\to\infty$ as well.
Thus, $p_n\to 0$ as claimed.
\qed

The bounds $a_{-}>17$ and $b<\squarter$ from \clm(CylinderBounds)
yield information about the spectrum of $D\buT(0)$,
using e.g.~the theorem below.
We note that the operator $D\buT(0)$ is compact,
for the reasons described before \clm(AntiRevContr).

\claim Theorem(EIG) {\rm([\rAKiKS])}
Let $A$ be a compact linear operator on a real Banach space $\real\times W$.
For $u\in\real$ and $w\in W$ write $A(u+w)=u'+w'$
with $u'\in\real$ and $w'\in W$.
Assume that there exist positive real numbers $b<a$
such that $\|w'\|\le b\max\{|u|,\|w\|\}$,
and such that $|u'|\ge a|u|$ whenever $|u|\ge\|w\|$.
Then $A$ has a simple eigenvalue of modulus $\ge a$
and no other eigenvalue of modulus $>b$.

Here, in the real setting,
a non-real number $\xi+i\eta$ is said to be an eigenvalue of $A$
if there exists nonzero vectors $x$ and $y$ such that
$Ax=\xi x-\eta y$ and $Ay=\xi y+\eta x$.

As a consequence of \clm(CylinderBounds) and \clm(EIG)
we have the following.

\claim Corollary(BuFHyperbolic)
The derivative $D\buT(p_\ast)$ at $p_\ast=0$
has a real eigenvalue $\lambda\ge a_{-}$
and no other spectrum outside the disk $|z|\le b$.
The local unstable manifold of $\buT$ at the fixed point $p_\ast=0$
is the graph of a real analytic function ${\tt p}_\ast$
from an open neighborhood of the origin in $\UU$ to $\WW$.
Furthermore, ${\tt p}_\ast$ extends
to a real analytic function on an open neighborhood of $[-h_0,h_0]$,
taking values in $\{w\in\WW: \|w\|<r\}$.

The existence and real analyticity
of the local unstable manifold near $p_\ast=0$
follows from standard theorems on invariant manifolds.
Its extension is obtained by iterating $\buT$
and using \clm(CylUniq).

Clearly \clm(BuFHyperbolic) translates trivially
to an analogous result for the transformation $\buR_3^2$.
In fact, an analogous result holds for $\buR_3$ as well,
since $P_\ast$ is a fixed point of $\buR_3$.
Our reason for considering the second iterate of $\buR_3$
in this section is that it was easier to find a good isomorphism $\LL$
in this case.

\subsection Proof of \clm(ProdConv), Part I

Our goal here is to prove \clm(ProdConv),
based on on estimates that can be (and have been) verified
with the aid of a computer.
A transversality condition that is needed,
and the claim that $s_\ast=0$, will be proved in Section 6.

Here we consider the unstable manifold of $\buR_3$
at $P_\ast$ to be a curve in the cylinder $C_0'=P_\ast+\LL C_0$ rather than a graph.
A possible parametrization of this curve is given by
${\tt P}_\ast(t)=P_\ast+\LL(tu_0+{\tt p}_\ast(t))$,
with $t$ ranging in $[-h_0,h_0]$.
Here $u_0\in\UU$ is the unit vector mentioned earlier.
The projection of ${\tt P}_\ast$ onto $\LL\UU$
is a strictly increasing function;
and as $t$ increases from $-h_0$ to $h_0$,
the curve ${\tt P}$ connects the bottom of $C_0'$ to the top.

Given a real number $c>0$, consider the extension of $\buR_3$
to one-parameter families of pairs $s\mapsto{\tt P}(s)$,
defined by the equation
$$
\buF_c({\tt P})(s)=\buR_3({\tt P}(cs))\,.
\equation(SimpleFamRG)
$$

\claim Lemma(AMIterates)
Consider the \AM family ${\tt P}$ for $\lambda=e^s$.
Then there exist real numbers $\sigma,\eps>0$ and an integer $m>0$,
such that the following holds.
Consider the curve ${\tt P}_0=\buF_{\alpha^3}^{2m}({\tt P})$.
Then ${\tt P}_0(s)$ lies in the interior of $C_0'$ for $-\sigma<s<\sigma$.
As $s$ is increased from $-\sigma-\eps$ to $\sigma+\eps$,
the curve ${\tt P}_0$ enters the cylinder $C_0'$
through the bottom (corresponding to $u=-h_0$ for $C_0$)
at $s=-\sigma$
and leaves it through the top (corresponding to $u=-h_0$ for $C_0$),
at $s=\sigma$.
An analogous statement holds if
${\tt P}$ is replaced by $\buF_{\alpha^3}({\tt P})$.

Our (computer-assisted) proof of this lemma
implements the transformation $\buF_{\alpha^3}$
on a space of curves $s\mapsto{\tt P}(s)$ in $\FF_\rho$
that are analytic in a disk $|s|<\delta$ of radius $\delta=2^{-96}$.
In this space, we determine bounds on the curve
${\tt P}_0=\buF_{\alpha^3}^{2m}({\tt P})$ for $m=66$
that imply the claims of \clm(ProdConv) via strict inequalities.
For further details we refer to Section 8.

\smallskip
Some immediate consequences of \clm(AMIterates) are the following.
There exists an increasing sequence $n\mapsto s_n^{-}$
and a decreasing sequence $n\mapsto s_n^{+}$, with $s_n^{-}<s_n^{+}$ for all $n$,
such that the curve ${\tt P}_n=\buF_1^{2n}({\tt P}_0)$
enters the bottom of the cylinder $C_0'$
at the parameter value $s_n^{-}$
and leaves the top of the cylinder for the first time at a parameter value $s_n^{+}$.

Pick a parameter value $s_\infty$ that belongs to
$[s_n^{-},s_n^{+}]$ for every $n$.
Then the orbit $n\mapsto\buR_3^{2n}({\tt P}_0(s_\infty))$
converges to $P_\ast$ by \clm(CylUniq).
If $n$ is sufficiently large,
then the pair ${\tt P}_n(s_\infty)$ is close enough
to $P_\ast$ for perturbative arguments to apply.
In particular, if ${\tt P}_n$ intersects the local stable manifold transversally,
then we can use the graph transform [\rHPS] to characterize convergence.

The graph transform $\buF$ associated with $\buR_3^2$ takes the form
$\buF({\tt P})=\buR_3^2\circ{\tt P}\circ R$.
Here, $R=R({\tt P})$ is a real analytic function defined near $s_\infty$.
Its dependence on ${\tt P}$ is real analytic
and can be chosen in such a way that $\buF$ has an attracting fixed point.
By construction, this fixed point is the (canonically parametrized)
local unstable manifold of $\buR_3$ at $P_\ast$.
If ${\tt P}$ is any curve in the domain of $\buF$,
then the sequence $k\mapsto\buF^k({\tt P})$ converges to the fixed point of $\buF$,
and $k\mapsto R\bigl(\buF^k({\tt P})\bigr)$ converges to the function $s\mapsto\mu_2^{-1}s$.
In fact, $R$ can be chosen affine,
at the expense of possibly weakening the rate of convergence.

\smallskip
Assume now that $s_\infty$ must have the value $0$,
and that the curves ${\tt P}_n$, for $n$ sufficiently large,
are transversal to the local stable manifold of $\buR_3$ at $P_\ast$.
These properties will be proved in Section 6.

In this case, the re-parametrization function $R$ can be chosen linear.
Since convergence of $k\mapsto R\bigl(\buF^k({\tt P}_n)\bigr)$
to the function $s\mapsto\mu_2^{-2}s$ is exponential,
we can in fact choose the fixed re-parametrization $s\mapsto\mu_2^{-2}s$ at each step.
This show that, if ${\tt P}$ is the \AM family with parameter $\lambda=e^s$,
then the sequence $n\mapsto\buF_{\mu_2^{-2}}^n({\tt P}_0)$
converges to ${\tt P}_\ast$, modulo a one-time linear re-parametrization.
Here, we assume that the local unstable manifold ${\tt P}_\ast$
has been parametrized in such a way that it is a fixed point of $\buF_{\mu_2^{-2}}$.

The same holds if ${\tt P}$ is replaced by $\buF_{\alpha^3}({\tt P})$.
So the above arguments can be repeated
for the graph transform associated with $\buR_3$.
We note that the pairs ${\tt P}_\ast(t)$ are limits
of renormalized \AM pairs, so they are commuting.

Since transversality to the local stable manifold
is stable under small perturbations,
we can repeat the same arguments for a sufficiently
small perturbation of the \AM curve.
The only difference is that a one-time affine re-parametrization
is needed.

\section The Lyapunov exponent

The Lyapunov exponent will be used
to establish a connection between observable quantities
and local properties of $\buR_3$ near the fixed point $P_\ast$.

\subsection The critical coupling

Let $G=(\alpha,A)$ be the anti-reversible \AM map
with coupling constant $\lambda=e^s$.
Let $\lambda_\ast$ be a value of the parameter $\lambda$
for which the pair $P=(F,G)$ with $F=(1,\idmat)$
gets attracted to $P_\ast$ under the iteration of $\buR_3^2$.
The goal here is to show that $\lambda_\ast=1$.

First, we claim that $\lambda_\ast\le 1$.
To see why, consider $\lambda>1$.
Then the Lyapunov exponent $L(G)=\log\lambda$ is positive.
Thus, by Proposition 6.4 in Subsection 6.2,
the sequence of functions $n\mapsto\|A^{\ast q_n}(\alpha^n\bdot)\|$
cannot stay bounded on $[-1,1]$ as $n\to\infty$.
So we cannot have $\buR^n(P)\to P_\ast$ as $n\to\infty$ along multiples of $3$.

Our next goal is to exclude the possibility $\lambda_\ast<1$.
The following is Theorem 3.4 in [\rAvAC].
It applies to the \AM map $G=(\alpha,A)$, for any irrational $\alpha$
whose continued fraction denominators $q_n$ satisfy $\lim_n q_n^{-1}\log q_{n+1}=0$.

\claim Theorem(AqPowerBound)  {\rm([\rAvAC])}
Let $0<\lambda<1$ and assume that $E$ belongs to the spectrum of $H_\lambda^\alpha$.
There exists constants $a,b,c>0$ such that for all $q>0$,
$$
\|A^{\ast q}(z)\|\le bq^a\,,\qquad |\Im z|\le c\,.
\equation(AqPowerBound)
$$

An analogous result for Diophantine $\alpha$ was probably proved earlier.
Corollary 4.5 in [\rAvJi] comes close,
but it considers only the real domain.

In what follows, if $G=(\alpha,A)$ is an arbitrary skew-product map,
we will write $A^{\alpha\ast q}$ instead of $A^{\ast q}$
for the product \equ(Aastqx), in order to emphasize the dependence on $\alpha$.

Let $G$ be an anti-reversible \AM map
for energy zero and $\alpha$ the inverse golden mean.
Consider the corresponding pair $P=(F,G)$ with $F=(1,\idmat)$,
and its \RG iterates $P_n=\buR_3^{2n}$.
We choose here even powers of $\buR_3$, so that the reflections
$L(0)$ that are part of the scaling $L_3=L(0)e^{-\vartheta J}$ cancel.
And for the fixed point $P_\ast$, the rotations cancel as well,
since $L(0)e^{-\vartheta J}L(0)=e^{-\vartheta J}$.
Thus, in order to simplify notation, consider $\buR_3$ with $L_3=\idmat$.
Then we can perform an initial rotation $A\mapsto e^{\vartheta J}Ae^{-\vartheta J}$
in such a way that $P_n\to P_\ast$ with $L_3=\idmat$ fixed.
Then $P_n=(F_n,G_n)$ with $F_n=(1,B_n)$ and $G_n=(\alpha,A_n)$, where
$$
A_n(x)=A^{\alpha\ast q_{6n}}\bigl(\alpha^{6n}x\bigr)\,,\qquad
B_n(x)=A^{\alpha\ast q_{6n-1}}\bigl(\alpha^{6n}x\bigr)\,.
\equation(AnBnDef)
$$
As described earlier, the factors $A_\ast$ and $B_\ast$
of the fixed point $P_\ast$ are entire,
due to the analyticity-improving property of $\buR_3$.
For the same reason, we have convergence $A_n\to A_\ast$
and $B_n\to B_\ast$, uniformly on compact subsets of $\complex$.

Let now $u$ and $v$ be fixed but arbitrary nonnegative integers, not both zero.
Then
$$
A_n^{\alpha\ast u}(\,\bdot+v)B_n^{1\ast v}
\to A_\ast^{\alpha\ast u}(\,\bdot+v)B_\ast^{1\ast v}\,,
\equation(ABConvOne)
$$
uniformly on compact subsets of $\complex$.

\claim Proposition(PPisPoly)
Assume that $\lambda_\ast<1$.
Then $x\mapsto A_\infty^{\alpha\ast u}(x+v)B_\infty^{1\ast v}(x)$
is a polynomial whose degree cannot be larger
than the constant $a$ in \clm(AqPowerBound).

\proof
Let $M$ be some fixed $2\times 2$ matrix and define
$$
f_n=\tr\bigl(M A_n^{\alpha\ast u}(\,\bdot+v)B_n^{1\ast v}\bigr)\,,\quad
f_\ast=\tr\bigl(M A_\ast^{\alpha\ast u}(\,\bdot+v)B_\ast^{1\ast v}\bigr)\,.
\equation(fnxfxDef)
$$
By \clm(AqPowerBound) we have a bound
$$
|f_n(z)|\le 2b(uq_{6n-1}+vq_{6n})^a
\le C(\alpha u+v)^a\alpha^{-6na}\,,\qquad
|\Im z|\le c\alpha^{-6n}\,,
\equation(fAqPowerBound)
$$
for some fixed constant $C>0$.
Now restrict to a disk $|z|\le r$ of radius $r>0$.
If $n$ is sufficiently large, then
$$
\partial_z^k f_n(z)={k!\over 2\pi i}\int_{\Gamma_n}
{f_n(\zeta)\over(\zeta-z)^{k+1}}\,d\zeta\,,
\equation(fnCauchy)
$$
where $\Gamma_n$ is the path along the two
circles in $\complex/\bigl(\alpha^{-6n}\integer\bigr)$ at $\Im\zeta=\pm c\alpha^{-6n}$.
Using the bound \equ(fAqPowerBound)
and the fact that $|\Gamma_n|=2\alpha^{-6n}$, we have
$$
\eqalign{
\bigl|\partial_z^k f_n(z)\bigr|
&\le{k!\over 2\pi}2\alpha^{-6n}\bigl(c\alpha^{-6n}-r\bigr)^{-k-1}
C(\alpha u+v)^a\alpha^{-6na}\cr
&\le C_k(\alpha u+v)^a\alpha^{6n(k-a)}\,,\cr}
\equation(fnCauchyBound)
$$
for some constant $C_k>0$.
Thus, if $k>a$, then the derivative $\partial_z^k f_\ast(z)=\lim_n\partial_z^k f_n(z)$
vanishes on the disk $|z|<r$.

This shows that $f_\ast$ is a polynomial of degree $\lfloor a\rfloor$ or less.
Since $M$ was arbitrary, we conclude that
$x\mapsto A_\infty^{\alpha\ast u}(x+v)B_\infty^{1\ast v}(x)$
is a polynomial of degree $\lfloor a\rfloor$ or less.
\qed

\claim Theorem(PPisConst) $\lambda_\ast=1$.

\proof
We have already established that $\lambda_\ast\le 1$.
Assume for contradiction that $\lambda_\ast<1$.
In what follows, $x$ and $y$ denote real numbers.
Given any $m>0$, denote by $d_m$
the polynomial degree of $A_\ast^{\ast m}$,
meaning the maximal degree of any of the components of $A_\ast^{\ast m}$.
Clearly, $x\mapsto A_\ast^{\alpha\ast m}(\omega x+z)$
has degree $d_m$ as well, for any complex numbers $\omega\ne 0$ and $z$.

Consider now the identity \equ(asEvenPowerAoiy),
which holds whenever $(\alpha,A)$ is anti-reversible.
By taking the trace of $(-1)^m A_\circ^{\ast 2m}(iy)$,
we obtain the square of the Hilbert-Schmidt norm of
$U(iy)=A_\circ((m-\sfrac{1}{2})\alpha+iy)\cdots A_\circ(\sfrac{\alpha}{2}+iy)$.
Applying this to $A=A_\ast$, we see that
the function $y\mapsto A^{\ast 2m}_\ast(iy)$ must have degree $2d_m$.
Iterating this argument shows that $A_\ast^{\ast 2^j}$ has degree $2^jd_1$.
But the degree of $A_\ast^{\ast 2^j m}$ cannot exceed $a$, by \clm(PPisPoly).
Thus, we must have $d_1=0$.

An analogous argument shows that $B_\ast$ has degree $0$ as well.
But we know that neither $A_\ast$ nor $B_\ast$ are constant.
So $\lambda_\ast\ge 1$.
\qed

\subsection A Lyapunov exponent for pairs

Let $\alpha$ be an irrational number between $0$ and $1$.
To a pair $P=(F,G)$ with $F=(1,B)$ and $G=(\alpha,A)$,
we associate a renormalized pair as in \equ(RGDef) by setting
$$
\buR(P)=\bigl(\Lambda^{-1}G\Lambda,\Lambda^{-1}FG^{-c}\Lambda\bigr)\,.
\equation(RGDefAgain)
$$
For simplicity, we restrict here to the trivial scaling
$\Lambda=\Lambda(P)$, given by $\Lambda(x,y)=(\alpha x,y)$.
Consider now the iterates $P_n=\buR^n(P)$.
The components of $P_n$ are of the form
$F_n=(1,B_n)$ and $G_n=(\alpha_n,A_n)$.
After choosing a suitable norm for the factors $A_n$,
a Lyapunov-type exponent for pairs can be defined by setting
$$
\ell(P)=\limsup_{n\to\infty}q_n^{-1}\log\|A_n\|\,.
\equation(ellP)
$$
where $q_n$ is the $n$-th continued fraction denominator for $\alpha$.

Consider first the case where $F=(1,\idmat)$.
Then the functions $A_n$ are scaled versions of $A^{q_n}$.
To be more precise, $A_n(x)=A^{\ast q_n}(\bar\alpha_n x)$,
where $\bar\alpha_n=\alpha_0\alpha_1\cdots\alpha_{n-1}$.
Taking the $\sup$-norm in \equ(ellP) on a domain $|x|<\eps$,
we have $\ell(P)=\ell_\eps(G)$, where
$$
\ell_\epsilon(G)=\limsup_{n\to\infty}
q_n^{-1}\log\sup_{|x|\le\epsilon}\|A^{\ast q_n}(\bar\alpha_n x)\|\,.
\equation(ellG)
$$
Here, and in what what follows, we assume that $A$
is a continuous $1$-periodic function on $\real$, taking values in $\rmSL(2,\real)$.
Notice that $\ell_\epsilon(G)\le L(G)$ by Furman's theorem [\rFurman].

\claim Proposition(RGLyap)
Let $\alpha\in\real\setminus\rational$ be of finite type,
in the sense that the sequence $n\mapsto\alpha_n$ is bounded away from zero.
Then the $\limsup$ in \equ(ellG) exists as a limit,
and $\ell_\epsilon(G)$ agrees with the Lyapunov exponent $L(G)$.

\proof
We may assume that $L(G)>0$.
Then, by Oseledets' theorem  [\rOseledets], we have
$$
\lim_q{1\over q}\log{\|A^{\ast q}(x_0)v_0\|\over\|v_0\|}=L(G)\,,
\equation(nosupFixOne)
$$
for almost every $x_0\in\real$, and for all vectors
$v_0$ outside some one-dimensional subspace of $\real^2$
that can depend on $x_0$.
Now fix such an $x_0$ and $v_0$.

Using the three-gap theorem [\rSur,\rSos,\rSwi],
we can find an integer $k>0$ and sequences $n\mapsto t_n$ and $n\mapsto s_n$
of positive integers, such that
$$
q_n\le t_n\le q_{n+k}\,,\qquad
|x_0+t_n\alpha-s_n|\le\epsilon\bar\alpha_n\,.
\equation(nosupFixOne)
$$
Setting $x_n=x_0+t_n\alpha-s_n$
and $v_n=A^{\ast t_n}(x_0)v_0$, we have
$$
\eqalign{
{1\over q_n}\log{\|A^{\ast q_n}(x_n)v_n\|\over\|v_n\|}
&=\left(1+{t_n\over q_n}\right){1\over q_n+t_n}\log{\|A^{\ast(q_n+t_n)}(x)v_0\|\over\|v_0\|}\cr
&\quad-{t_n\over q_n}\;{1\over t_n}\log{\|A^{\ast t_n}(x)v_0\|\over\|v_0\|}\,.\cr}
\equation(nosupFixOne)
$$
Notice that $1\le t_n/q_n\le C$ for some fixed constant $C$.
Thus, the right hand side of \equ(nosupFixOne)
converges to $L(G)$ as $n\to\infty$.
\qed

In what follows, we assume that $\alpha$ is the inverse golden mean.
Then $c=1$ in the equation \equ(RGDefAgain), and $\alpha_n=\alpha$ for all $n$.
In addition, we restrict to pairs in the set $\FF_\rho'$ defined below,
where $\rho=(\rho_\ssF,\rho_\ssG)$ is assumed to satisfy
$$
\thalf\alpha^{-2}<\rhoG\,,\qquad
\thalf\alpha+\alpha\rhoG\le\rhoF\le\alpha^{-1}\rhoG\,.
\equation(RROneDomainCond)
$$

\claim Definition(FFrhoPrime)
Define $\FF_\rho'$ to be the set of pairs $P=(F,G)$
in $\FF_\rho^{\tinyskip r}$ with the property
that the maps $F$ and $G$ commute,
and that their factors have determinant $1$.

The condition \equ(RROneDomainCond)
guarantees that $\buR$ defines a dynamical system on $\FF_\rho'$.
This condition is satisfied e.g.~for the values
$\rho=(2,\sfrac{11}{8})$ and $\rho=(3,2)$ that are used in \clm(AntiRevContr)
and \clm(RevContr), respectively.

Using that $q_n^{-1}=\bigl(1+\alpha^2\bigr)\alpha^n+\OO\bigl(\alpha^{3n}\bigr)$,
we can rewrite \equ(ellP) as
$$
\ell(P)=\limsup_{n\to\infty}L_n(P)\,,
\equation(ellPDef)
$$
where
$$
L_n(P)=\bigl(1+\alpha^2\bigr)\alpha^n\log\|A_{n,\circ}\|_\rhoG\,,\qquad
A_{n,\circ}(x)=A^{\ast q_n}_\circ\bigl(\alpha^n x\bigr)\,.
\equation(LnPDef)
$$
Here, we have used that scaling and ``symmetrizing'' commute,
as mentioned after \equ(TildeSymmHatSymm).

\demo Remark(buRiiiLyap)
If we restrict $n$ in \equ(LnPDef) to multiples of $3$
and replace $\buR^3$ by $\buR_3$,
then the weaker domain condition \equ(RRThreeDomainCond) is sufficient.
In addition, the restriction to commuting pairs can be omitted in this case.
We will not do this here, simply to avoid complicating notation.

Our main reason for considering a Lyapunov exponent
based on the functionals $L_n$ is that
it transforms conveniently under renormalization:
$$
L_n\bigl(\buR(P))=\alpha^{-1}L_{n-1}(P)\,,\qquad n=1,2,\ldots
\equation(LnRP)
$$
So we have $\ell\bigl(\buR(P))=\alpha^{-1}\ell(P)$.
Clearly, $\ell$ vanishes on any periodic orbit of $\buR$.

\claim Proposition(ellVersusL)
Let $G=(\alpha,A)$ be an anti-reversible \AM map with $\alpha$
the inverse golden mean.
Then the limit $\displaystyle\lim_{n\to\infty}L_n(P)$ exists
and is equal to $L(G)$.

\proof
Let $r=\rho_\ssG$.
Consider the $\sup$ norm in \equ(ellG) for small $\epsilon>0$.
Since the evaluation map $g\mapsto g(x)$
is continuous on $\GG_r$ for $|x|\le r$,
this norm is bounded from above by $C\|A_{n,\circ}\|_r$
for some fixed constant $C>0$.
So we have $\displaystyle\liminf_{n\to\infty}L_n(P)\ge L(G)$.

Next, consider the \AM map $G_\delta$
with factor $A_\delta(x)=A(x+i\delta)$.
Using the (well known) fact that $\log\bigl\|A^{\ast q}(x+i\delta)\bigr\|$
is a convex function of $\delta$, we have
$$
L(P)\le
\limsup_{n\to\infty}q_n^{-1}
\sup_{|x|\le\epsilon}\log\bigl\|A^{\ast q_n}_\circ\bigl(\alpha^n x\pm i\delta\bigr)\bigr\|
\le\ell_\epsilon(G_\delta)=L(G_\delta)\,,
\equation(ellPvsLGdelta)
$$
for $\epsilon>0$ sufficiently large (depending only on $r$)
and every $\delta>0$.
The ``$\pm$'' in the equation \equ(ellPvsLGdelta) includes a maximum over the two signs.
But by anti-reversibility, the supremum over $|x|\le\epsilon$
does not depend on the sign.

Now we can use the fact [\rAvG] that
$L(G_\delta)=\max\bigl\{0,\log\lambda+2\pi\delta\,\bigr\}$.
Thus, taking $\delta\to 0$ in \equ(ellPvsLGdelta) yields
$\displaystyle\limsup_{n\to\infty}L_n(P)\le L(G)$.
This proves the claim in \clm(ellVersusL).
\qed

For more general pairs in $\FF_\rho'$,
the $\limsup$ in \equ(ellPDef) may not be a limit.
An easy way to cure this problem is to define
a slightly different Lyapunov-type exponent as follows:
$$
L(P)=\limsup_{n\to\infty}{1\over 1+\alpha^2}
\Bigl[L_n(P)+\alpha^2 L_{n-1}(P)\Bigr]\,.
\equation(LnToL)
$$

\claim Theorem(LnToL)
Assume that $P\in\FF_\rho'$ for some choice
of $\rho=(\rhoF,\rhoG)$ that satisfies \equ(RROneDomainCond).
Then the sequence $n\mapsto L_n(P)+\alpha^2 L_{n-1}(P)$ is decreasing.
As a consequence, the $\limsup$ in \equ(LnToL) is achieved as a limit,
and $L$ is upper semi-continuous on $\FF_\rho'$.
Furthermore,
the value of $L(P)$ does not depend on the choice of $\rho$.

\proof
For $n\ge 0$ define
$$
b_n=b_n(\rho)=\log\|B_{n,\circ}\|_\rhoF\,,\qquad
a_n=a_n(\rho)=\log\|A_{n,\circ}\|_\rhoG\,.
\equation(anbnDef)
$$
Assume now that $n\ge 1$.
Then the domain conditions \equ(RROneDomainCond) guarantee that
$$
b_n\le a_{n-1}\,,\qquad a_n\le b_{n-1}+a_{n-1}\,.
\equation(anbnIter)
$$
Using that $1+\alpha=\alpha^{-1}$, this yields the bound
$$
\eqalign{
L_n(P)+\alpha^2L_{n-1}(P)
&=\bigl(1+\alpha^2\bigr)
\bigl[\alpha^n a_n+\alpha^{n+1}a_{n-1}\bigr]\cr
&\le\bigl(1+\alpha^2\bigr)
\bigl[\alpha^n b_{n-1}+\alpha^n a_{n-1}+\alpha^{n+1}a_{n-1}\bigr]\cr
&\le\bigl(1+\alpha^2\bigr)
\bigl[\alpha^n a_{n-2}+\alpha^n(1+\alpha)a_{n-1}\bigr]\cr
&=\alpha^2L_{n-2}(P)+L_{n-1}(P)\,,\cr}
\equation(LPairMonotone)
$$
for $n\ge 2$. This shows that the sequence
$n\mapsto L_n(P)+\alpha^2L_{n-1}(P)$
is decreasing and thus has a limit.
Since each $L_k$ is continuous, the limit is upper semi-continuous.

Next consider another domain parameter $\varrho$ satisfying \equ(RROneDomainCond).
Let us write $L_n(\varrho)$ instead of $L_n(P)$ when the domain parameter is $\varrho$,
and $L_n(\rho)$ when the domain parameter is $\rho$.
Using that $\buR$ is analyticity-improving,
there exists $k>0$ and a constant $c_k>0$ such that
$$
\|B_{n+k}\|_\varrhoF\le e^{c_k}\|B_n\|_\rhoF^{q_{n-2}}\|A_n\|_\rhoG^{q_{n-1}}\,,\quad
\|A_{n+k}\|_\varrhoF\le e^{c_k}\|B_n\|_\rhoF^{q_{n-1}}\|A_n\|_\rhoG^{q_n}\,,
\equation(BnkAnkBound)
$$
for every $n\ge 0$.
Taking logarithms and using the identity
$$
\twomat{q_{k-2}}{q_{k-1}}{q_{k-1}}{q_k}\twovec{\alpha}{1}
=\twomat{0}{1}{1}{1}^k\twovec{\alpha}{1}
=\alpha^{-k}\twovec{\alpha}{1}\,,
\equation(preLrhoCompOne)
$$
we find that
$$
\alpha^2L_{n+k-1}(\varrho)+L_{n+k}(\varrho)
\le C_k\alpha^n+\alpha^2L_{n-1}(\rho)+L_n(\rho)\,,
\equation(LrhoCompOne)
$$
with $C_k=\bigl(1+\alpha^2\bigr)^2\alpha^kc_k$.
An analogous inequality holds if $\varrho$ and $\rho$ are exchanged.
This shows that $L(P)$ is independent of the choice of $\rho$.
\qed

Using upper semicontinuity and \equ(LnRP),
together with the fact that the pair $P_\ast$ commutes,
we immediately obtain the following.

\claim Corollary(WsLyapZero)
Consider a pair $P\in\FF_\rho'$. Then $L(\buR(P))=\alpha^{-1}L(P)$.
So in particular, $L(P_\ast)=0$.
Furthermore, if $\buR^{3n}(P)\to P_\ast$ as $n\to\infty$, then $L(P)=0$.

\subsection Transversal intersection

At this point we can complete the proof of \clmno(ProdConv).
We already know from Subsection 6.1 that $s_\ast=1$.
What remains to be proved is that some \RG iterate
of the \AM family intersects the local stable manifold $\WW^s$
of $\buR_3$ transversally, and that $\mu_2\ge\alpha^{-3}$.

To this end, consider the splitting $\FF_\rho=W^u\oplus W^s$,
where $W^u$ is the unstable subspace for the operator $D\buR_3(P_\ast)$
and $W^s$ the stable subspace.
Then $P_\ast+W^s$ is tangent to $\WW^s$ at $P_\ast$.
And $\WW^s$ is the graph of a real analytic function from $W^s$ to $P_\ast+W^u$.
To be more precise, this holds locally, near $P_\ast$.
So we restrict our analysis to a suitable open ball $B$
in $\FF_\rho$ that is centered at $P_\ast$.
Using that $\WW^s$ is a graph,
we can define the height of a pair $P\in B$ relative to $\WW^s$
in the direction of $W^u$.

Consider the \AM family $s\mapsto{\tt P}(s)$ associated with the factor \equ(SinAMo)
for $\lambda=e^s$. Define ${\tt P}_n=\buF_1^n({\tt P})$,
where $\buF_1$ denotes the pointwise version of $\buR_3$ defined by \equ(SimpleFamRG).
Notice that $\buR_3$ may be replaced by $\buR^3$, since the \AM pairs commute.
Recall that Lyapunov exponent of ${\tt P}(s)$ is $\max\{0,s\}$.
So by \equ(RGLyap), the Lyapunov exponent of ${\tt P}_n(s)$ is $\alpha^{-3n}s$.
And recall from Subsection 6.1 that $0$ is the unique value of $s$
for which ${\tt P}(s)$ is attracted to $P_\ast$ under the iteration of $\buR_3$.
If $n$ is sufficiently large, so that the pair ${\tt P}_n(0)$ lies on $\WW^s$ in $B$,
then the pairs ${\tt P}_n(s)$ for $s>0$ cannot lie on $\WW^s$.
Here, we have \clm(WsLyapZero).
In what follows, $k$ denotes some (large) value of $n$ for which this holds.

Now choose $t>0$ such that ${\tt P}_\ast(t)$ belongs to $B$.
Let $\Sigma_0$ be a codimension $1$ subspace of $\FF_\rho$
that passes through ${\tt P}_\ast(t)$ and is transversal
to the unstable manifold ${\tt P}_\ast$.
We may assume also that $\Sigma_0$ does not intersect $\WW^s$.
For $n=1,2,\ldots$, define $\Sigma_n$ to be the inverse image
of $\Sigma_{n-1}$ under $\buR_3$, restricted to $B$.
If $B$ has been chosen sufficiently small,
then the $\lambda$-Lemma [\rPalisMelo] guarantees the following.
The sets $\Sigma_n$ are real analytic manifolds.
Furthermore, the sequence $n\mapsto\Sigma_n$ accumulates
at the stable manifold $\WW^s$ of $\buR_3$ at $P_\ast$
asymptotically like $\mu_2^{-n}$.
To be more precise, if ${\tt Q}$ is any (real analytic) curve in $B$
that crosses $\WW^s$ transversally at the point ${\tt Q}(0)$,
then near this point, and for sufficiently large $k$, the curve ${\tt Q}$
intersects $\Sigma_n$ for a unique parameter value $s_n$.
Furthermore, the sequence $n\mapsto\mu_2^ns_n$ converges to a nonzero constant.

Consider now the curve ${\tt Q}={\tt P}_k$.
Denote by $h(s)$ the height of ${\tt Q}(s)$ above $\WW^s$
in the direction of $W^u$.
By analyticity, we have $h(s)=as^m+\OO\bigr(s^{m+1}\bigr)$ for some $m\ge 1$, with $a\ne 0$.
Thus, for sufficiently large $n$,
the curve ${\tt Q}$ intersects $\Sigma_n$ at some parameter value $s_n>0$.
We may choose the smallest such value.
By construction, the point $\buR^{3n}({\tt Q}(s_n))$ lies on $\Sigma_0$.
So we must have $\mu_2^n s_n^m\ge b$ for sufficiently large $n$,
where $b$ is some positive constant.
Using that the Lyapunov exponent is bounded on $B$,
and that $\buR^{3n}({\tt Q}(s_n))$ has Lyapunov exponent $\alpha^{-3(n+k)}s_n$,
we also have $\alpha^{-3n}s_n\le c$ for some $c>0$.
Thus, $\mu_2\ge\alpha^{-3m}$.
We know from \clm(CylinderBounds) that $\mu_2\le\sqrt{19}$.
And for $m\ge 2$ we have $\alpha^{-3m}>17$.
So we must have $m=1$.
This implies that ${\tt Q}=\buF_1^k({\tt P})$
intersects $\WW^s$ transversally, and that $\mu_2\ge\alpha^{-3}$.
This completes the proof of \clm(ProdConv).

\demo Remark(WuPosLyap)
The parameter values $s_n$ depend on the value of $k$
defining the curve ${\tt Q}={\tt P}_k$.
Writing $s_n=s_{k,n}$ we obtain $\buR^{3(k+n)}({\tt P}(s_{k,n}))\to{\tt P}_\ast(t)$
in the limit $k\to\infty$.
If we assume that $\mu_2=\alpha^{-3}$,
then this implies that ${\tt P}_\ast(t)$ has a positive Lyapunov exponent.

\section Supercritical maps

The main goal in this section is to prove
Theorems \clmno(SuperCritFix) and \clmno(AMZeros).

\subsection Limiting zeros

Here we consider the factor-normalization
that was mentioned before \clm(SuperCritFix).
Recall from \equ(FullMapRG) that $\buR_3=\buN\circ\RR_3\circ\buC$,
where $\buC$ is a commutator correction,
and where $\buN$ normalizes determinants to $1$.
The factor-normalization of a pair
$P=((1,B),(\alpha,A))$ is component-wise,
$\buN(P)=((1,\MM(B),(\alpha,\NN(A))$, with $\MM$ and $\NN$ of the form
$$
\MM(B)=M(B)^{-1}B\,,\qquad\NN(A)=N(A)^{-1}A\,.
\equation(SuperCritMN)
$$
Up to now we have used
$M(B)=\det(B)^{1/2}$ and $N(A)=\det(A)^{1/2}$.

Here, we are interested only in pairs that commute
and whose factors have constant nonnegative determinants.
So we omit the commutator correction $\buC$
and choose for $M$ and $N$ the norms in $\FF_\rho$.
Other choices would work equally well,
as long as they guarantee that the orbits of $\buR_3$ remain bounded
without tending to zero.
In order to simplify the description,
we also choose a trivial $y$-scaling $L_3=\idmat$.

Let $K_0$ be a set of pairs $P\in\FF_\rho$
whose norm is bounded by some fixed constant,
and that satisfy $L(P)\ge\eps$ for some fixed $\eps>0$.
Recall that the transformation $\buR^3$ is compact,
as described before \clm(AntiRevContr).
Here, and in what follows, we assume that the domain parameter $\rho$
satisfies the condition \equ(RRThreeDomainCond).
Consequently, the sets $K_n=\buR^{3n}(K_0)$ for $n>0$ all have compact closures.
Denote by $K_\ast$ the set of all accumulation points
from the sequence $n\mapsto K_n$.
This set is compact and invariant under $\buR_3$.
By taking $K_0$ invariant under conjugacies by a rotation,
the limit set $K_\ast$ has the same property.
Notice that the pairs in $K_\ast$ belong to $\FF_\rho$,
so their factors are analytic.

Let $P\in K_0$ and define $P_n=\buR_3^n(P)$ for $n\ge 0$.
By \equ(MatGFiG), \equ(MatGiFGiFGi), and \equ(TildeSymmHatSymm),
the symmetric factors associated with the pairs $P_n$
are related via
$$
\eqalign{
N_n B_{n,\circ}(x)
&=
A_{n-1,\circ}\bigl(\alpha^3\bigl(x-\tfrac{\alpha^{-1}}{2}\bigr)\bigr)
B_{n-1,\circ}\bigl(\alpha^3x\bigr)^\dagger
A_{n-1,\circ}\bigl(\alpha^3\bigl(x+\tfrac{\alpha^{-1}}{2}\bigr)\bigr)\,,\cr}
\equation(LLtildeB)
$$
and
$$
\eqalign{
M_n A_{n,\circ}(x)
&=
A_{n-1,\circ}\bigl(\alpha^3\bigl(x+\alpha^{-1}\bigr)\bigr)^\dagger
B_{n-1,\circ}\bigl(\alpha^3\bigl(x+\tfrac{\alpha^{-1}}{2}\bigr)\bigr)\times\cr
&\quad\times
A_{n-1,\circ}\bigl(\alpha^3x\bigr)^\dagger
B_{n-1,\circ}\bigl(\alpha^3\bigl(x-\tfrac{\alpha^{-1}}{2}\bigr)\bigr)
A_{n-1,\circ}\bigl(\alpha^3\bigl(x-\alpha^{-1}\bigr)\bigr)^\dagger\,.\cr}
\equation(LLtildeA)
$$
Here, $M_n$ and $N_n$ are the normalization factors
that appear in the definition of $P_n=\buR_3(P_{n-1})$.
Assume now that $K_0$ is a set of pairs $P=(F,G)$
with $F=(1,\idmat)$ and $G=(\alpha,A)$,
and with the property that $A_\circ(0)=-J$.
The latter condition is satisfied e.g.~for the anti-reversible \AM family.
In this case, \equ(asOddPowerAoZero) implies that
$$
A_{n,\circ}(0)=\pm\eps_n J\,,\qquad\eps_n=\OO\bigl(\alpha^{-3n}\bigr)\,.
\equation(AnoepsBound)
$$
Here, $\eps_n$ is a product of $n$ factors $N_m^{-1}$ or $M_m^{-1}$
for $m=1,2,\ldots,n$.
The given estimate on $\eps_n$ uses the fact that,
under the iteration of $\RR_3$,
the norms grow at least as quickly as the Lyapunov exponents.
Thus, as $n\to\infty$, the values at $x=0$
of the factors $A_{n,\circ}$ tend to zero, uniformly in our sequence $n\mapsto K_n$.

There are many other values of $x$ where $A_{n,\circ}(x)$ tends to zero.
An example is $x=\alpha^{-1}$.
At this value of $x$, the last factor in \equ(LLtildeA)
is of size $\OO\bigl(\alpha^{-3n}\bigr)$ for large $n$.
Thus, as $n\to\infty$, the values at $x=\alpha^{-1}$
of the factors $A_{n,\circ}$ tend to zero as well.
The search for such zeros can be made more systematic as follows.

Let $P=((1,B),(\alpha,A))$ be a pair in $K_\ast$,
and let $\tilde P=\bigl(\bigl(1,\tilde B\bigr),\bigl(\alpha,\tilde A\bigr)\bigr)$
be the image of $P$ under the transformation $\buR_3$.
Denote by $\AA$ and $\BB$ be the set of zeros of $A_\circ$ and $B_\circ$, respectively.
Then \equ(LLtildeB) shows that the set of zeros of $\tilde B_\circ$
includes the set
$$
\tilde\BB=\biggl[\alpha^{-3}\BB
\bigcup_{m=\pm1}\Bigl(\alpha^{-3}\AA+\tfrac{m}{2}\alpha^{-1}\Bigr)\biggr]\,,
\equation(LLtildeBB)
$$
and \equ(LLtildeA) shows that the set of zeros of $\tilde A_\circ$
includes
$$
\tilde\AA=\biggl[\,\bigcup_{m=\pm1}\Bigl(\alpha^{-3}\BB+\tfrac{m}{2}\alpha^{-1}\Bigr)
\bigcup_{m=0,\pm1}\Bigl(\alpha^{-3}\AA+m\alpha^{-1}\Bigr)\biggr]\,.
\equation(LLtildeAA)
$$
This defines a transformation that
maps a pair of sets of complex numbers $\PP=(\BB,\AA)$
to a pair $\tilde\PP=\bigl(\tilde\BB,\tilde\AA\bigr)$.
Consider iterates $\PP_1,\PP_2,\ldots$
under this transformation $\PP\mapsto\tilde\PP$,
starting with a pair $\PP_0=(\BB_0,\AA_0)$.

Notice that the imaginary parts of non-real points
in $\BB_0$ or $\AA_0$ expand by a factor $\alpha^{-3}$
under the transformation $\PP\mapsto\tilde\PP$.
Thus, the limit sets ($\limsup$ or $\liminf$)
$\BB_\ast$ and $\AA_\ast$ are always subsets of $\real$.

In the case at hand, $\AA_0$ includes $0$, since $A_\circ(0)=0$ for all pairs in $K$.
So consider $\BB_0=\{\}$ and $\AA_0=\{0\}$.
Then $\AA_1$ includes $0$ as well, due to the map $x\mapsto\alpha^{-3}x$
that appears in \equ(LLtildeAA) for $m=0$.
So it is clear that $\BB_{n+1}\supset\BB_n$ and $\AA_{n+1}\supset\AA_n$ for all $n$.
Thus, $\BB_n\nearrow\BB_\infty$ and $\AA_n\nearrow\AA_\infty$
for a pair of sets $\PP_\infty=(\BB_\infty,\AA_\infty)$.

A zero that appears for the \AM family
but that is not included in the above set $\AA_\infty$ is $x=\shalf$.
And $x=-\shalf$ appears as well, due to anti-reversibility.
To see how these zeros occur,
denote by $C_n(x)$ the product of the last two factors in \equ(LLtildeA).
An explicit computation shows that
$$
C_n(x)
=A^{\ast q_{3n-2}}_0\bigl(\alpha^{3n}x+\tfrac{q_{3n-1}}{2}\alpha\bigr)\,.
\equation(LLhalfThree)
$$
Using that $q_k\alpha-q_{k-1}=(-1)^k\alpha^{k+1}$,
one finds that $C_n(\sfrac{\sigma}{2})=J$ for $\sigma=(-1)^{3n-1}$.
This implies that the factor $A_\circ$ for an anti-reversible
\AM pair $P\in K_\ast$ has a zero at $x=\pm\shalf$.
The same holds for other anti-reversible pairs
$P=((1,\idmat),(\alpha,A))$ with $A$ of Schr\"odinger type \equ(SchrA).

To see how the zeros at $\pm\shalf$
propagate under iteration of $\PP\mapsto\tilde\PP$,
consider $\BB_0=\{\}$ and $\AA_0=\{-\sfrac{1}{2},\sfrac{1}{2}\bigr\}$.
Then $\AA_1$ includes $\pm\shalf$ as well,
since $\pm\shalf$ is a fixed point of the map
$x\mapsto\alpha^{-3}\mp\alpha^{-1}$ that appear in \equ(LLtildeAA).
Again we have $\BB_n\nearrow\BB_\infty$ and $\AA_n\nearrow\AA_\infty$
for a pair of sets $\PP_\infty=(\BB_\infty,\AA_\infty)$.

\subsection The supercritical fixed point

The recursion relations \equ(LLtildeBB) and \equ(LLtildeAA)
can be obtained more easily from pairs of commuting skew-product maps
with factors that take values in $\rmGL(1,\real)$ instead of $\rmGL(2,\real)$.
A map $g:(x,y)\mapsto(x+\alpha,a(x)y)$ of this type
will again be written as $g=(\alpha,a)$.
Given a pair $p=(f,g)$ of such maps $f=(-1,b)$ and $g=(\alpha,a)$,
the renormalized pair $\RR(p)$ is given by
$$
\RR(p)=\bigl(\lambda^{-1}g\lambda\,\bcomma\,\lambda^{-1}fg^c\lambda\bigr)\,,\qquad
\lambda(x,y)=(-\alpha x,y)\,,
\equation(trigRG)
$$
with the same positive integer $c$ as in \equ(RGDef).
The analogue of anti-reversibility here is the requirement
that $b_\circ$ be even and $a_\circ$ odd.
The analogue of a Schr\"odinger pair
is a pair $p=(f,g)$ with components
$f=(-1,1)$ and $g=(\alpha,a)$, where $a$ is periodic with period $1$.

Let us restrict now to the case where $\alpha$ is the inverse golden mean.
Then $c=1$ at every \RG step. If $p$ is anti-reversible,
then the relation between the zeros for $p=(f,g)$
and the zeros for $\tilde p=\RR^3(p)$
is trivially given by \equ(LLtildeBB) and \equ(LLtildeAA).

\smallskip
Consider again skew product maps with factors in $\rmGL(2,\real)$.
Let $P_0$ be an anti-reversible \AM pair with $\lambda>1$.
Consider the pairs $P_n=\buR_3^n(P_0)$ for $n\ge 1$
and the associated symmetric factors $B_{n,\circ}$ and $A_{n,\circ}$.
These factor can be obtained iteratively via \equ(LLtildeB) and \equ(LLtildeA).
The following are numerical observations.

\demo Observation(ObsOne)
As $n$ increases, the factors $B_{n,\circ}$ and $A_{n,\circ}$
approach symmetric matrices with constant null spaces.

In other words, if we start with $K_0=\{P\}$,
then the limit set $K_\ast$ appears to consist of pairs
whose symmetric factors are of the form
$$
B_\circ(x)=b_\circ(x)WW^\transpose\,,\qquad
A_\circ(x)=a_\circ(x)VV^\transpose\,,
\equation(antirevBoAo)
$$
where $W$ and $V$ are constant unit vectors in $\real^2$.
Notice that $b_\circ=\tr(B_\circ)$ has to be even
and $a_\circ=\tr(A_\circ)$ odd.
This follows from the (anti)reversibility property \equ(antirevBoAo).
The part of \dem(ObsOne) that we cannot prove
is that $A_\circ$ is symmetric, and that $V$ and $W$ are constant.
It is possible to give some formal arguments,
but we will not do this here.
If $V$ and $W$ are assumed to be constant,
then the fact that $F$ has to commute with both $G$ and $G^\dagger$
implies that $W=V^\dagger$.

\demo Observation(ObsTwo)
The sequence of pairs
$n\mapsto P_n$ converges to a fixed point of $\buR_3$.

\smallskip
To be more precise, if we use the trivial $y$-scaling $L_2=\idmat$,
as we we have done so far in this section,
then convergence is to a period $2$ of $\buR_3$.
The traces $b_\circ$ and $a_\circ$ reproduce after one step of $\buR_3$,
but the directions $V$ and $W=V^\dagger$ only repeat after two steps.
Using instead $L_3=L(\vartheta)$ of the form \equ(Ltheta),
with $\vartheta=\vartheta(P)$ chosen appropriately,
the sequence $n\mapsto P_n$ converges numerically
to a fixed point of $\buR_3$, with the properties described in \clm(SuperCritFix).

\proofof(SuperCritFix)
We give only a sketch here,
since the proof follows closely the steps used in [\rKochTrig].

First, notice that the fixed point equation for a pair of the form
described in \clm(SuperCritFix) reduces to a fixed point equation
for the functions $b_\diamond$ and $a_\diamond$.
This is essentially the fixed point equation for the third power $\RR^3$
of the operator $\RR$ defined by \equ(trigRG), with $c=1$,
except for a constant re-normalization of the factors $b$ and $a$.

For simplicity, let us re-define $b_\diamond$ and $a_\diamond$
to be the symmetric factors of the desired fixed point.
Then we have good guess where their zeros are:
starting with $\BB_0=\{\}$ and $\AA_0=\{-\shalf,0,\shalf\}$,
an iteration of the map $\PP\mapsto\tilde\PP$
yields a limit pair $\PP_\infty=(\BB_\infty,\AA_\infty)$
that should be the set of zeros associated with the
symmetric factors of the fixed point.

For the actual construction, it is more convenient
to start with $\BB_0=\{\}$ and $\AA_0=\half\integer$.
This corresponds to starting with the pair $P$ whose symmetric factors are
$$
B_\circ(x)=\stwomat{0}{0}{0}{1}\,,\qquad
A_\circ(x)=2\sin(2\pi x)\stwomat{1}{0}{0}{0}\,,
\equation(lambdainfty)
$$
and then iterating $\buR^3$ with $L_3=\bigl[{0~1\atop 1~0}\bigr]$.
Notice that the symmetric factor $A_\circ$ in this equation
is the limit as $\lambda\to\infty$ of the anti-symmetric \AM factor
(in its standard form), after dividing it by $\lambda$.

The symmetric factors that are being generated
by iterating the map $(B_\circ,A_\circ)\mapsto\bigl(\hat B_\circ,\hat A_\circ\bigr)$
defined by \equ(MatGFiG) and \equ(MatGiFGiFGi)
have their zeros modulo $\shalf$ on the orbit of the point $x=0$
under the translation $x\mapsto x+\alpha$ on the circle $\torus/2=\real/(2\integer)$.
Thus, by the three-gap theorem [\rSur,\rSos,\rSwi],
the gaps between adjacent zeros on $\torus/2$ take at most $3$ distinct values.
To be more specific, there are exactly three distinct gaps,
except when the orbit has length $q_{3n}$, at which point the largest gap gets closed.
So after $n$ iterations,
the resulting factor $\hat B_\circ$ has $q_{3n-1}$ zeros on $\torus/2$
with $3$ distinct gaps,
while the factor $\hat A_\circ$ has $q_{3n}$ zeros on $\torus/2$
with $2$ distinct gaps.

The same applies to the iteration
$(B_\circ,A_\circ)\mapsto\bigl(\tilde B_\circ,\tilde A_\circ\bigr)$,
except that the circle gets enlarged by a factor $\alpha^{-3}$ at each step.
Furthermore, the zeros that lie within some fixed positive angle
of the origin reproduce after each step [\rKochTrig].
This yields the limiting sequences of zeros described in \clm(SuperCritFix).

The symmetric factors that are being generated during this iteration
have a Weierstrass representation as products,
determined (up to a constant factor) by their zeros.
Here, we use the fact that $b_\circ$ is even and $a_\circ$ odd,
so it is possible to work with products of order $\shalf$.
By controlling the zero sets $\BB_n$ and $\AA_n$,
one finds uniform convergence on compact sets
to a pair of limit functions $b_\diamond$ and $a_\diamond$.
For details we refer to [\rKochTrig], where \RG fixed points have been constructed
for skew-products with meromorphic factors.

The gap sizes listed in \equ(SuperCritFix)
were obtained by iterating the map $\PP\mapsto\tilde\PP$ a few times,
starting with the sets $\BB_0=\{\}$ and $\AA_0=\{-\shalf,0,\shalf\}$.
\qed

The following will be used in the proof of \clm(AMZeros) below.

\claim Proposition(MinimalAo)
Let $\BB_0=\{\}$. Let $X=\{0\}$ or $X=\{-\shalf,\shalf\}$
or $X=\{-\shalf,0,\shalf\}$.
Assume that $X\subset\AA_1$ whenever $X\subset\AA_0$.
Then $\AA_0=X$ lead to the same limit pair $\PP_\infty$
as $\AA_0=X+\integer$.

\proof
For $n\ge 0$ define $\torus_n=\real/\bigl(\alpha^{3n}\integer\bigr)$.
Consider first the case $\AA_0=X+\integer$.
By construction, the points in $\BB_n$ constitute $|X|$ orbits of length $q_{3n-1}$
for the translation $x\mapsto x+\alpha^{3n+1}$ on the circle $\torus_n$,
and the points in $\AA_n$ constitute $|X|$ orbits of length $q_{3n}$.
Furthermore, the orbits are symmetric with respect to $x\mapsto-x$.

Consider now $\AA_0=X$ instead of $\AA_0=X+\integer$.
Then we must get the same orbits on $\torus_n$, simply by counting points.
For $m>0$, denote by $a(n,m)$ and $b(n,m)$
the number of points in $\AA_n$ and $\BB_n$, respectively,
that lie within a distance $m$ from the origin in $\torus_n$.

Next, construct $\BB_n$ and $\AA_n$ as subsets of $\real$ instead of $\torus_n$.
Now the number of points in $\BB_n$ or $\AA_n$
that lie within a distance $m$ of the origin
can be smaller than $b(n,m)$ or $a(n,m)$, respectively.
However, this does not happen if $n$ is sufficiently large.
The reason is that map $\PP\mapsto\tilde\PP$ is ``expanding''
by a factor $\alpha^{-3}$.
To be more precise, let $c>\alpha^3$.
Then for $m$ larger than some constant that only depends on $c$,
the points in $\tilde\BB_n\cap[-m,m]$ and $\tilde\AA_n\cap[-m,m]$
are determined via $\PP\mapsto\tilde\PP$ from the points in
$\BB_n\cap[-cm,cm]$ and $\AA_n\cap[-cm,cm]$,
if $n$ is sufficiently large.
This shows that $\AA_0=X$ leads to the same limit pair $\PP_\infty$
as $\AA_0=X+\integer$.
\qed

Notice that $\liminf_n\BB_n=\limsup_n\BB_n$ and $\liminf_n\AA_n=\limsup_n\AA_n$
in the above.

\proofof(AMZeros)
Consider the choice $M(B)=\|B_\circ\|$ and $N(A)=\|A_\circ\|$
in the definition \equ(SuperCritMN) of our factor-normalization.
Let $P\in K_\ast$.
Given that $P$ belongs to $\FF_\rho$ and
lies in the range of $\buR_3^n$ for every $n$,
the factors $B$ and $A$ extend analytically to all of $\complex$.
Furthermore, $b_\circ=\tr(B_\circ)$ is even and $a_\circ=\tr(A_\circ)$ odd,
due to the anti-reversibility property \equ(antirevBoAo).

By construction, the symmetric factors $B_\circ$ and $A_\circ$ have norm $1$
and determinant $0$.
{}From \equ(asOddPowerAo) we see that $B_\circ(iy)$ or $-B_\circ(iy)$
is the limit of real positive matrices,
for every $y\in\real$ satisfying $|y|<\rho_\ssF$.
Here, we have used that $m=q_{3n-1}/2$ is an integer
whose parity is independent of $n$.
Since $B(x)$ is a symmetric matrix for imaginary $x$,
it is symmetric for all $x$.

The claim in \clm(AMZeros) concerning the zeros of $B$
and $A$ follows from our discussion in the previous subsection,
together with the characterization
of the set of zeros for the functions $b_\diamond$ and $a_\diamond$
given in the proof of \clm(SuperCritFix),
as well as \clm(MinimalAo).
\qed

\demo Remark(WuPdiamond)
If one assumes that some point $P$ on the unstable manifold
of $\buR_3$ at $P_\ast$ converges to $P_\diamond$ under iteration of $\buR_3$,
then it is possible to show that $\mu_2=\alpha^{-3}$.
The reason is that the asymptotic behavior
under renormalization becomes trivial in this case.

\section Computer estimates

What remains to be done is to verify the estimates in Lemmas
\clmno(AntiRevContr), \clmno(RevContr), \clmno(AAEigen),
\clmno(CylinderBounds), and \clmno(AMIterates).
This is carried out with the aid of a computer.
This part of the proof is written in the programming language Ada [\rAda]
and can be found in [\rFiles].
The following is meant to be a rough guide for the
reader who wishes to check the correctness of our programs.

\subsection Enclosures and data types

Bounds on a vector $x$ in a space $\XX$,
also referred to as enclosures for $x$,
are given here by sets $X\subset\XX$ that include $x$
and are representable as data on a computer.
Data of type {\tt Ball} are pairs ${\tt B}=({\tt B.C},{\tt B.R})$,
where {\tt B.C} and {\tt B.R} are representable numbers, with ${\tt B.R}\ge 0$.
In a Banach algebra $\XX$ with unit $\bfOne$,
the enclosure associated with a {\tt Ball} {\tt B}
is the ball ${\tt B}_\XX=\{x\in\XX:\|x-({\tt B.C})\bfOne\|\le{\tt B.R}\}$.
Our other enclosures are closed convex subsets of $\XX$
that admit a canonical decomposition
$$
S=\sum_n x_n{\tt B(n)}_\XX\,,
\equation(genEnlosure)
$$
where each $x_n$ is a representable element in $\XX$,
and where each {\tt B(n)} is a {\tt Ball} with center $\bfZero$ or $\bfOne$.
Notice that a {\tt Ball} can have radius zero.

Consider now a disk $D=\{z\in\complex:|z|<\rho\}$
with representable radius $\rho>0$.
An analytic function $g:D\to\XX$ admits a Taylor series representation
$g(z)=\sum_{n=0}^\infty g_nz^n$ with coefficients $g_n\in\XX$.
Denote by $\GG$ the space of all such functions
that have a finite norm $\|g\|=\sum_{n=0}^\infty\|g_n\|\rho^n$.
Assume that $\XX$ carries a type of enclosures named {\tt Scalar}.
To each such enclosure ${\tt S}_\XX\subset\XX$
we can associate a set ${\tt S}_\GG\subset\GG$
by replacing each ball ${\tt B(n)}_\XX$
in the decomposition \equ(genEnlosure) of $S={\tt S}_\XX$ by the ball ${\tt B(n)}_\GG$.

Given an integer $D>0$,
our enclosures for functions in $\GG$ are specified by data of type {\tt Taylor1}.
A {\tt Taylor1} is in essence a pair {\tt P=(P.F,P.C)},
where ${\tt P.F}\le D$ is a nonnegative integer,
and where {\tt P.C} is an {\tt array(0 .. D) of Scalar}.
The corresponding enclosure is the set
$$
{\tt P}_\GG=\sum_{n=0}^{m-1}{\tt P.C(n)}_\XX\PP_n
+\sum_{n=m}^D{\tt P.C(n)}_\GG\PP_n\,,\qquad\PP_n(z)=z^n\,,
\equation(TaylorEnc)
$$
where $m={\tt P.F}$.
Notice that the first sum in \equ(TaylorEnc) is a polynomial,
while each term in the second sum is in general a non-polynomial.
This allows for efficient estimates in the problem considered here.
For precise definitions we refer to the Ada package {\tt Taylors1}.

Quadruples of {\tt Taylor1} define a type {\tt TMat2}
that is used for enclosures of $2\times 2$ matrices $A$ with entries in $\GG$.
By including another component
$\alpha\in\integer\bigl[\half\sqrt{5}-\half\bigr]$
we obtain enclosures for skew-product maps $G=(\alpha,A)$.
The corresponding data type is named {\tt Skew}.
Enclosures on pairs $P=(F,G)$ are described by data of type {\tt Skew2}.
For details we refer to the child package {\tt Taylors1.Skews2}.

In most of our packages, the type {\tt Scalar} is generic,
meaning unspecified,
except for a list of available operations.
When instantiated with {\tt Scalar => Ball},
the packages {\tt Taylors1} and {\tt Taylors1.Skews2} define enclosures
for the spaces $\GG_\rho$, $\GG_\rho^4$, and $\FF_\rho$ described in Subsection 3.5.

A particular instantiation of {\tt Taylors1},
with {\tt Scalar => Ball}, is named {\tt S\_T} in the package {\tt FamRG}.
The resulting type {\tt Taylor1} is named {\tt TScalar}.
A second instantiation of {\tt Taylors1}, with {\tt Scalar => TScalar},
is named named {\tt T\_T}.
The type of {\tt Taylor1} defined by {\tt T\_T}
describes analytic functions from a disk $|s|<\delta$ to $\GG_\rho$.
Now it suffices to instantiate the child package {\tt T\_T.Skews2},
to obtain {\tt Skew2}-type enclosures on analytic curves in the space $\FF_\rho$.
This covers all major data types used in our programs.

\subsection Bounds and procedures

After having defined enclosures for (elements in) the various spaces
that are needed in our analysis,
we need to implement bounds on maps between these space.
In this context, a bound on a map $f:\XX\to\YY$
is a function $F$ that assigns to a set $X\subset\XX$
of a given type ({\tt Xtype}) a set $Y\subset\YY$
of a given type ({\tt Ytype}), in such a way that
$y=f(x)$ belongs to $Y$ whenever $x\in X$.
In Ada, such a bound $F$ can be implemented by defining
an appropriate {\tt procedure F(X\col in Xtype\scol Y\col out Ytype)}.
In practice, the domain of {\tt F} is restricted.
If {\tt X} does not belong to the domain of {\tt F},
the {\tt F} raises an {\tt Exception} which causes the program to abort.

Our type {\tt Ball} is defined in the package {\tt MPFR.Floats.Balls},
using centers {\tt B.C} of type {\tt MPFloat}
and  radii {\tt B.R} of type {\tt LLFloat}.
Data of type {\tt MPFloat} are high-precision floating point numbers,
and the elementary operations for this type are implemented
by using the open source MPFR library [\rMPFR].
Data of type {\tt LLFloat} are standard extended floating-point numbers [\rIEEE]
of the type commonly supported by hardware.
Both types support controlled rounding.
Bounds on the basic operations for this type {\tt Ball}
are defined and implemented in {\tt MPFR.Floats.Balls}.

Using the definition \equ(TaylorEnc) of a {\tt Taylor1}-type enclosure,
it is clearly possible to implement
a bound {\tt Prod} on the map $(f,g)\mapsto f*g$ from $\GG\times\GG$ to $\GG$.
This and other basic bounds that include the type {\tt Taylor1}
are defined in package {\tt Taylors1}.
Basic bounds that involve the types {\tt Skew} and {\tt Skew2}
are defined in the package {\tt Taylors1.Skew2}.
This includes a bound {\tt Normalize} on the map $\buN$
defined by \equ(DivSqrtDet).
It also includes a bound {\tt Inv} on the map $G\mapsto G^\dagger$
and a bound {\tt Prod\_GFG} on the product $(F,G)\mapsto GFG$.
Combining the two yields bounds on the products
that appear in the definition  of the operator $\buR_3$.

Bounds on problem-specific maps such as $\buR_3$
are mostly defined in child packages of {\tt Taylors1.Skew2}.
Among the exceptions are the bounds named {\tt Equalize}
on the normalization transformations defined in Subsections 3.2 and 4.1.
The package {\tt Taylors1.Skew2.RG3r} implements bounds
on operations such as $\buR_6$ that are used only for reversible pairs,
while {\tt Taylors1.Skew2.RG3a} implements bounds
that are specific to anti-reversible pairs.
This includes procedures named {\tt Commutize}
that provide bounds on the commutator-correction map $\buC$.
A bound on the derivative of $\buC$ is named {\tt DCommmutize}.
The same naming convention is used for other derivative bounds.

Up to this level,
the same bounds can be used for pairs and for families of pairs.
To choose one or the other,
is suffices to instantiate the package {\tt Taylors1}
and its children with the desired type of {\tt Scalar}.
But for derivative bounds on maps such as $\buM$,
we need to be able to enumerate the degrees of freedom,
so it matters whether {\tt Scalar} encloses a number or a Taylor series.

\subsection Linear operators and modes

In the packages {\tt MapR}, {\tt FamRG}, and their children,
degrees of freedom are associated with a type {\tt Mode}.
To simplify the discussion, consider first the space $\GG_\rho$.
In this case, a ``coefficient mode'' $c_n$ represents
a monomial $\PP_n$ of degree $n$,
and an ``error mode'' $e_n$ represents the unit ball
in the subspace of all functions $g=e\PP_n$ with $e\in\GG_\rho$.
So {\tt Taylor1}-type enclosure \equ(TaylorEnc)  in $\GG_\rho$
is a finite linear combination of modes for $\GG_\rho$.
A proper collection of normalized modes $\{h_1,h_2,\ldots,h_m\}$
defines an analogue of a finite basis for $\GG_\rho$.
Due to our choice of norm in $\GG_\rho$,
the operator norm of a bounded linear $T:\GG_\rho\to\GG_\rho$
is simply $\|T\|=\max_n\|Th_n\|$.
Given that each mode $h_n$ admits a representation of type {\tt Taylor1},
a bound on $\|T\|$ is easily obtained from a bound on $T$.

This generalizes readily to the space $\FF_\rho^{\tinyskip r}$.
The corresponding type {\tt CMode} is defined in the package {\tt MapRG}.
For a ``basis'' of such modes we use the type {\tt CModes}.
In {\tt MapRG} we also instantiate two generic packages
{\tt Linear} and {\tt Linear.Contr}
that implement bounds on a quasi-Newton map (and its derivative)
of the type \equ(ContrRGThree)
in terms of bounds on the given map (and its derivative).
Using these facilities,
the child package {\tt MapRG.RG3a} implements a bound
{\tt DContrNorm} on the norm of $D\buM$,
where $\buM$ is the transformation defined by \equ(ContrRGThree).
This procedure is used to verify the bound on
the operator norm of $D\buM(p)$ in \clm(AntiRevContr).

The operator $M$ that enters the definition \equ(ContrRGThree)
is a ``matrix'' on a subspace spanned by finitely many coefficient modes,
and on the complementary subspace it is the zero operator.
This matrix is included in [\rFiles] as a data file {\tt Contr3aMat.132}.
The matrix $M$ that is used in the definition \equ(ContrRGSix)
is included in [\rFiles] as a file {\tt Contr3rMat.134}.
The operator $\LL$ that enters the definition \equ(buFDef)
is a ``matrix'' $L$ on a subspace spanned by finitely many coefficient modes,
and on the complementary subspace it is the identity operator.
This matrix $L$ is included in the file {\tt Iso.132.28}.
A bound on its inverse is obtained and saved by running the program {\tt Invert\_Iso}.

The child package {\tt MapRG.RG3r} is an analogue of {\tt MapRG.RG3a},
but for reversible pairs.

The package {\tt FamRG} and its children
define an analogue of the {\tt MapRG} hierarchy,
but for analytic curves $s\mapsto{\tt P}(s)$.
The modes for this space are named {\tt DModes}.
However, these modes are not needed in our current proof.
(An earlier version implemented the graph transform method
that is described in Subsection 5.2.)
The child package {\tt FamRG.RG3a} implements the bounds
that are used in our proof of \clm(AMIterates).
This includes a bound {\tt Plain\_FamRG3}
on the transformation $\buF_c$ defined in \equ(SimpleFamRG),
for the special value $c=\alpha^3$.

\subsection Organizing the bounds

Our proof of \clm(AMIterates) is organized
in the main program {\tt Iter\_AM\_Fam\_Sin6}.
Starting with an enclosure for the anti-reversible \AM family
for $\lambda$ near $1$,
it does little more than iterating the above-mentioned bound {\tt Plain\_FamRG3}.
The bounds are in essence numerical computations,
but, as should be clear by now,
they include rigorous estimates of truncation errors and rounding errors.

The same is true for our proof of the remaining lemmas.
The approximate solutions $\bar P$ referred to
in Lemmas \clmno(AntiRevContr) and \clmno(RevContr)
have been computed numerically beforehand.
First we used a numerical versions of {\tt Iter\_AM\_Fam\_Sin6}
(or {\tt Iter\_AM\_Fam\_Cos6} in the reversible case)
to obtain rough approximate fixed points for $\buR_3^2$.
Then the approximations were improved via the procedures {\tt IterContr}
in the packages {\tt MapRG.RG3a} and {\tt MapRG.RG3r}, respectively.
(Numerical versions of our programs are obtained simply
by using for {\tt Scalar} the type {\tt Rep} instead of {\tt Ball}.)
The results are in the data files
{\tt approx-Fix3a.trunc} and {\tt approx-Fix6r.trunc} in [\rFiles].

The main programs that are used to verify the estimates in Lemmas
\clmno(AntiRevContr), \clmno(RevContr), and \clmno(CylinderBounds)
are {\tt Check\_\-RG3a\_Fixpt}, {\tt Check\_RG6r\_Fixpt}, and {\tt CheckNorms\_DRGN3}.
They do little more than instantiating the required packages
with the appropriate parameters, reading data files if needed,
and then handing the task over to the proper procedure(s)
in the instantiated packages.

Our programs were run successfully on
a standard desktop machine, using a public version of the gcc/gnat compiler [\rGnat].
Instructions on how to compile and run these programs can be found in the file
{\tt README} that is included with the source code in [\rFiles].

\bigskip
\references

{\ninepoint

\item{[\rHarp]} P.G.~Harper,
{\sl Single band motion of conduction electrons in a uniform magnetic field},
Proc. Phys. Soc. Lond. A {\bf 68}, 874--892 (1955).

\item{[\rSos]} V.T.~S\'os,
{\sl On the distribution {\rm mod $1$} of the sequence $n\alpha$},
Ann. Univ. Sci. Budapest,
E\"otv\"os Sect. Math. {\bf 1}, 127--134 (1958).

\item{[\rSur]} J.~Sur\'anyi,
{\sl \"Uber die Anordnung der Vielfachen einer reelen Zahl {\rm mod $1$}},
Ann. Univ. Sci. Budapest,
E\"otv\"os Sect. Math. {\bf 1}, 107--111 (1958).

\item{[\rSwi]} S.~\'Swierczkowski,
{\sl On successive settings of an arc on the circumference of a circle},
Fundamenta Mathematicae {\bf 46}, 187--189 (1959).

\item{[\rHof]} D.R.~Hofstadter,
{\sl Energy levels and wave functions of Bloch electrons
in rational and irrational magnetic fields},
Phys. Rev. B {\bf 14}, 2239--2249 (1976).

\item{[\rHPS]} M.W.~Hirsch, C.C.~Pugh, M.~Shub,
{\sl Invariant Manifolds.}
Lecture Notes in Mathematics, Volume {\bf 583},
Springer Verlag, Berlin $\cdot$ Heidelberg $\cdot$ New York (1977).

\item{[\rFei]} M.J.~Feigenbaum,
{\sl Quantitative universality for a class of non-linear transformations},
J. Stat. Phys. {\bf 19}, 25--52 (1978).

\item{[\rCouTr]} P.~Coullet, C.~Tresser,
{\sl It\'eration d'endomorphismes et groupe de renormalisation},
J. Phys. Colloque C {\bf 539}, C5--25 (1978).

\item{[\rKada]} L.P.~Kadanoff,
{\sl Scaling for a critical Kolmogorov--Arnold--Moser trajectory.}
Phys. Rev. Lett. {\bf 47}, 1641--1643 (1981).

\item{[\rTKNdN]} D.J.~Thouless, M.~Kohmoto, M.P.~Nightingale, M.~den Nijs,
{\sl Quantized Hall conductance in a two-dimensional periodic potential},
Phys. Rev. Lett. {\bf 49}, 405--408 (1982).

\item{[\rMcK]} R.S.~MacKay,
{\sl Renormalisation in Area Preserving Maps.}
Thesis, Princeton (1982).
World Scientific, London (1993).

\item{[\rBeSi]} J.~Bellissard, B.~Simon,
{\sl Cantor spectrum for the almost Mathieu equation},
J. Funct. Anal. {\bf 48}, 408--419 (1982).

\item{[\rJoMo]} R.~Johnson, J.~Moser,
{\sl The rotation number for almost periodic potentials},
Commun. Math. Phys. {\bf 84}, 403--438 (1982).

\item{[\rPalisMelo]} J.~Palis, W.~de~Melo,
{\it Geometric Theory of Dynamical Systems. An Introduction.}
Springer-Verlag, Berlin $\cdot$ New York (1982).

\item{[\rORJS]} S.~Ostlund, D.~Rand, J.~Sethna, E.~Siggia,
{\sl Universal properties of the transition
from quasiperiodicity to chaos in dissipative systems},
Physica, {\bf 8D}, 303-342 (1983).

\item{[\rAvSi]} J.~Avron, B.~Simon,
{\sl Almost periodic Schr\"odinger operators. II.
The integrated density of states},
Duke Math. J. {\bf 50}, 369--391 (1983).

\item{[\rEKW]} J.-P. Eckmann, H.~Koch, P.~Wittwer,
{\sl A computer-assisted proof of universality for area-preserving maps},
Mem. Amer. Math. Soc., Vol. {\bf 47}, No. 289, 1--121 (1984).

\item{[\rEckWit]} J.-P.~Eckmann, P.~Wittwer,
{\sl A complete proof of the Feigenbaum conjectures},
J. Stat. Phys. {\bf 46}, 455--475 (1987).

\item{[\rLanCC]} O.E.~Lanford III,
{\sl Renormalization group method for critical circle mappings},
Nonlinear evolution and chaotic phenomena,
NATO adv. Sci. Inst. Ser. B {\bf 176}, Plenum, New York, 25--36 (1988).

\item{[\rDeFaria]} E.~de Faria,
{\sl Proof of universality for critical circle mappings},
Thesis, CUNY (1992).

\item{[\rEliass]} L.~Eliasson,
{\sl Floquet solutions for the 1-dimensional quasi-periodic Schrödinger equation},
Comm. Math. Phys. {\bf 146}, 447--482 (1992).

\item{[\rSullivan]} D.~Sullivan,
{\sl Bounds, quadratic differentials and renormalization conjectures},
AMS Centennial Publications 2: Mathematics into twenty-first century (1992).

\item{[\rWieZa]} P.B.~Wiegmann, A.V.~Zabrodin,
{\sl Quantum group and magnetic translations.
Bethe ansatz solution for the Harper's equation},
Modern Phys. Lett. B {\bf 8}, 311--318 (1994).

\item{[\rLastii]} Y.~Last,
{\sl Zero measure spectrum for the almost Mathieu operator},
Comm. Math. Phys. {\bf 164}, 421--432 (1994).

\item{[\rMcMu]} C.~McMullen,
{\sl Renormalization and 3-manifolds which fiber over the circle},
Ann. Math. Studies {\bf 135}, Princeton University Press (1994).

\item{[\rBES]} J.~Bellissard, A.~van Elst, H.~Schulz-Baldes,
{\sl The Non-Commutative Geometry of the Quantum Hall Effect},
J. Math. Phys. {\bf 35}, 5373--5451 (1994).

\item{[\rFaKa]} L.D.~Faddeev, R.M.~Kashaev,
{\sl Generalized Bethe ansatz equations for Hofstadter problem},
Commun. Math. Phys. {\bf 169}, 181--191 (1995).

\item{[\rHKW]} Y.~Hatsugai, M.~Kohmoto, Y.-S.~Wu,
{\sl Quantum group, Bethe ansatz equations, and Bloch wave functions in magnetic fields},
Phys. Rev. B {\bf 53}, 9697--9712 (1996).

\item{[\rFurman]} A.~Furman,
{\sl On the multiplicative ergodic theorem for uniquely ergodic systems},
Ann. Inst. Henri Poincar\'e {\bf 33}, 797--815 (1997).

\item{[\rGJLS]} A.Y.~Gordon, S.~Jitomirskaya, Y.~Last, B.~Simon,
{\sl Duality and singular continuous spectrum in the almost Mathieu equation},
Acta Math. {\bf 178}, 169--183 (1997).

\item{[\rRuPi]} A.~R\"udinger, F.~Pi\'echon,
{\sl Hofstadter rules and generalized dimensions of the spectrum of Harper's equation},
J. Phys. A {\bf 30}, 117--128 (1997).

\item{[\rLyub]} M.~Lyubich,
{\sl Feigenbaum-Coullet-Tresser Universality and Milnor's Hairiness Conjecture},
Ann. Math. {\bf 149}, 319--420 (1999).

\item{[\rJito]} S.~Jitomirskaya,
{\sl Metal-insulator transition for the almost Mathieu operator},
Ann. of Math. {\bf 150}, 1159--1175 (1999).

\item{[\rMOW]} B.D.~Mestel, A.H.~Osbaldestin, B.~Winn,
{\sl Golden mean renormalization for a generalized Harper equation: The strong
coupling fixed point},
J. Math. Phys. {\sl 41} 8304--8330 (2000).

\item{[\rOsAv]} D.~Osadchy, J.E.~Avron,
{\sl Hofstadter butterfly as quantum phase diagram},
J. Math. Phys. {\bf 42}, 5665--5671 (2001).

\item{[\rYampol]} M~Yampolsky,
{\sl Hyperbolicity of renormalization of critical circle maps},
Publ. Math. Inst. Hautes \'Etudes Sci. {\bf 96}, 1--41 (2002).

\item{[\rdCLM]} A.~De Carvalho, M.~Lyubich, M.~Martens,
{\sl Renormalization in the H\'enon Family, I: Universality But Non-Rigidity},
J. Stat. Phys. {\bf 121}, 611--669 (2005).

\item{[\rDama]} D.~Damanik,
{\sl The spectrum of the almost Mathieu operator},
Lecture series in the CRC 701 (2008).

\item{[\rGoSch]} M. Goldstein and W. Schlag,
{\sl Fine properties of the integrated density of states
and a quantitative separation property of the Dirichlet eigenvalues},
Geom. Funct. Anal. {\bf 18}, 755--869 (2008).

\item{[\rAvAC]} A.~Avila,
{\sl The absolutely continuous spectrum of the almost Mathieu operator},
Preprint 2008.

\item{[\rOseledets]} V.~Oseledets,
{\sl Oseledets theorem},
Scholarpedia {\bf 3}, 1846 (2008).

\item{[\rGaJo]} D.~Gaidashev, T.~Johnson,
{\sl Dynamics of the universal area-preserving map associated with period doubling:
hyperbolic sets},
Nonlinearity {\bf 22}, 2487--2520 (2009).

\item{[\rAvJii]} A.~Avila, S.~Jitomirskaya,
{\sl The ten martini problem},
Ann. Math. {\bf 170}, 303--342 (2009).

\item{[\rAvJi]} A.~Avila, S.~Jitomirskaya,
{\sl Almost localization and almost reducibility},
J. Eur. Math. Soc. {\bf 12}, 93--131 (2010).

\item{[\rAKiKS]} G.~Arioli and H.~Koch,
{\sl Integration of dissipative PDEs: a case study},
SIAM J. Appl. Dyn. Syst. {\bf 9}, 1119--1133 (2010).

\item{[\rAKcRG]} G.~Arioli and H.~Koch,
{\sl The critical renormalization fixed point
for commuting pairs of area-preserving maps},
Comm ~Math. Phys. {\bf 295}, 415--429 (2010).

\item{[\rAvKrii]} A.~Avila, R.~Krikorian,
{\sl Monotonic cocycles},
Invent. Math. {\bf 202}, 271--331 (2015).

\item{[\rAvG]} A.~Avila,
{\sl Global theory of one-frequency Schr\"odinger operators},
Acta Math. {\bf 215}, 1--54 (2015).

\item{[\rSatij]} I.I.~Satija,
{\sl A tale of two fractals:
the Hofstadter butterfly and the integral Apollonian gaskets},
Eur. Phys. J. Spec. Top. {\bf 225}, 2533--2547 (2016)

\item{[\rKochAP]} H.~Koch,
{\sl On hyperbolicity in the renormalization of
near-critical area-preserving maps},
Discrete Contin. Dyn. Syst. {\bf 36}, 7029--7056 (2016).

\item{[\rKochAM]} H.~Koch,
{\sl Golden mean renormalization
for the almost Mathieu operator and related skew products},
J. Math. Phy. (to appear).

\item{[\rKKi]} H.~Koch, S.~Koci\'c,
{\sl Renormalization and universality of the Hofstadter spectrum},
Nonlinearity {\bf 33}, 4381--4389 (2020).

\item{[\rSatWil]} I.I~Satija, M.~Wilkinson,
{\sl Nests and chains of Hofstadter butterflies},
J. Phys. A: Math. Theor. {\bf 53}, 085703 (2020).

\item{[\rKochTrig]} H.~Koch,
{\sl On trigonometric skew-products over irrational circle-rotations},
Discrete Contin. Dyn. Syst. (to appear).

\item{[\rFiles]} H.~Koch.
The source code for our programs, and data files, are available at\hfill\break
\pdfclink{0 0 1}{{\tt web.ma.utexas.edu/users/koch/papers/skewunivers/}}
{http://web.ma.utexas.edu/users/koch/papers/skewunivers/}

\item{[\rAda]} Ada Reference Manual, ISO/IEC 8652:2012(E),
available e.g. at\hfil\break
\pdfclink{0 0 1}{{\tt www.ada-auth.org/arm.html}}
{http://www.ada-auth.org/arm.html}

\item{[\rGnat]}
A free-software compiler for the Ada programming language,
which is part of the GNU Compiler Collection; see
\pdfclink{0 0 1}{{\tt gnu.org/software/gnat/}}{http://gnu.org/software/gnat/}

\item{[\rIEEE]} The Institute of Electrical and Electronics Engineers, Inc.,
{\sl IEEE Standard for Binary Float\-ing--Point Arithmetic},
ANSI/IEEE Std 754--2008.

\item{[\rMPFR]} The MPFR library for multiple-precision floating-point computations
with correct rounding; see
\pdfclink{0 0 1}{{\tt www.mpfr.org/}}{http://www.mpfr.org/}

}

\bye